\documentclass[prb,amsmath,amssymb,amsbsy,twocolumn,showpacs]{revtex4-1}

\usepackage{amsfonts} 
\usepackage{amsmath}
\usepackage{amssymb}
\usepackage{cancel} 
\usepackage{bm}
\usepackage[usenames, dvipsnames]{color} 
\usepackage{dcolumn}
\usepackage{epic} 
\usepackage{epsfig}
\usepackage{eucal} 
\usepackage{feynmp}    
\usepackage[T1]{fontenc}
\usepackage{graphicx}
\usepackage[breaklinks,colorlinks = true,linkcolor = red,urlcolor=cyan,citecolor=red]{hyperref}
\usepackage{mathrsfs}
\usepackage{setspace}
\usepackage{soul}
\usepackage{subfigure}
\usepackage{times}
\usepackage{wrapfig} 
\usepackage{xy} 
\usepackage{wasysym} 

\def\a{\alpha}

\def\s{\sigma}
\def\t{\tau}

\definecolor{grey}{rgb}{0.4,0.4,0.5}
\definecolor{darkgreen}{rgb}{0,0.5,0}
\definecolor{darkred}{rgb}{0.6,0.0,0}
\definecolor{lightbrown}{rgb}{1,0.9,0.8}
\definecolor{brown}{rgb}{0.6,0.3,0.3}
\definecolor{darkblue}{rgb}{0,0,0.8}
\definecolor{darkmagenta}{rgb}{0.5,0,0.5}

\def\({\left(}
\def\){\right)}

\def\la{\label}

\def\be{\begin{equation}}
\def\ee{\end{equation}}
\def\ben{\begin{equation*}}
\def\een{\end{equation*}}

\newcommand{\bea}{\be\begin{aligned}}
\newcommand{\eea}{\end{aligned}\ee}
\newcommand{\bean}{\ben\begin{aligned}}
\newcommand{\eean}{\end{aligned}\een}

\newcommand{\bei}{\begin{itemize}}
\newcommand{\eei}{\end{itemize}}

\newcommand{\bee}{\begin{enumerate}}
\newcommand{\eee}{\end{enumerate}}

\newcommand{\bem}{\left (\begin{matrix}}
\newcommand{\eem}{\end{matrix} \right )}

\newcommand{\hF}{\mathcal{F}_{\hexagon}}
\newcommand{\lF}{\mathcal{F}_{\bigtriangleup}}
\newcommand{\rF}{\mathcal{F}_{\bigtriangledown}}

\begin{document}


\title{Phases and phase transitions of a perturbed Kekul\'{e}-Kitaev model}

\author{Eoin Quinn}
\email{epquinn@pks.mpg.de}
\affiliation{Max-Planck-Institut f\"ur Physik komplexer Systeme, N\"othnitzer Str. 38, 01187 Dresden, Germany}
\author{Subhro Bhattacharjee}
\affiliation{Max-Planck-Institut f\"ur Physik komplexer Systeme, N\"othnitzer Str. 38, 01187 Dresden, Germany}
\author{Roderich Moessner}
\affiliation{Max-Planck-Institut f\"ur Physik komplexer Systeme, N\"othnitzer Str. 38, 01187 Dresden, Germany}


\begin{abstract}
We study the quantum spin liquid phase in a variant of the Kitaev model where the bonds of the honeycomb lattice are distributed in a Kekul\'{e} pattern. The system supports gapped and gapless $Z_2$ quantum spin liquids with interesting differences from the original Kitaev model, the most notable being a gapped $Z_2$ spin liquid on a Kagome lattice. Perturbing the exactly solvable model with antiferromagnetic Heisenberg perturbations, we find a magnetically ordered phase stabilized by a quantum `order by disorder' mechanism, as well as an exotic continuous quantum phase transition between the topological spin liquid and this magnetically ordered phase. Using a combination of field theory and Monte-Carlo simulations, we find that the transition likely belongs to the $3D$-$XY\times Z_2$ universality class. 
\end{abstract}
\date{\today}
\maketitle
\section{Introduction}

Quantum spin liquids (QSL) represent prototypical condensed matter phases whose description requires understanding beyond the paradigm of spontaneous symmetry breaking.\cite{anderson1973resonating,PhysRevLett.86.1881,anderson1987resonating,wen2002quantum,balents2010spin,lee2008physics,lee2006doping} A QSL is a quantum paramagnet that can support quasiparticle excitations carrying quantum numbers which are fractions of the underlying microscopic degrees of freedom and hence are fundamentally different from random single spin flips of a thermal paramagnet or spin waves in a magnetically ordered state.\cite{anderson1987resonating,wen2002quantum,balents2010spin,lee2008physics,PhysRevLett.59.2095,lee2006doping} Systematic understanding of such phases, and phase transitions involving them, form an important area of current research in condensed matter physics. 

An important development in the understanding of QSLs came with the advent of exactly solvable spin Hamiltonians where the ground state is a QSL and low energy excitations are indeed fractionalized. Following the pioneering work of Kitaev,\cite{kitaev2006anyons,kitaev2003fault} several such models are now known \cite{PhysRevLett.90.016803,PhysRevB.79.024426,yao2007exact,yao2009algebraic,baskaran2009exact,PhysRevLett.105.067207,chua2011exact} and their investigations have enhanced our understanding of QSL phases. These Kitaev models usually do not have spin rotation symmetry and the suggestion that some of them may be realized in 5d transition metal compounds (like Iridates), due to the presence of strong spin-orbit coupling, has lead to a plethora of interesting studies regarding their properties.\cite{nussinov2013compass,PhysRevLett.105.027204,hermanns2014quantum,lee2014heisenberg}

\begin{figure}
\centering
\includegraphics[scale=0.3]{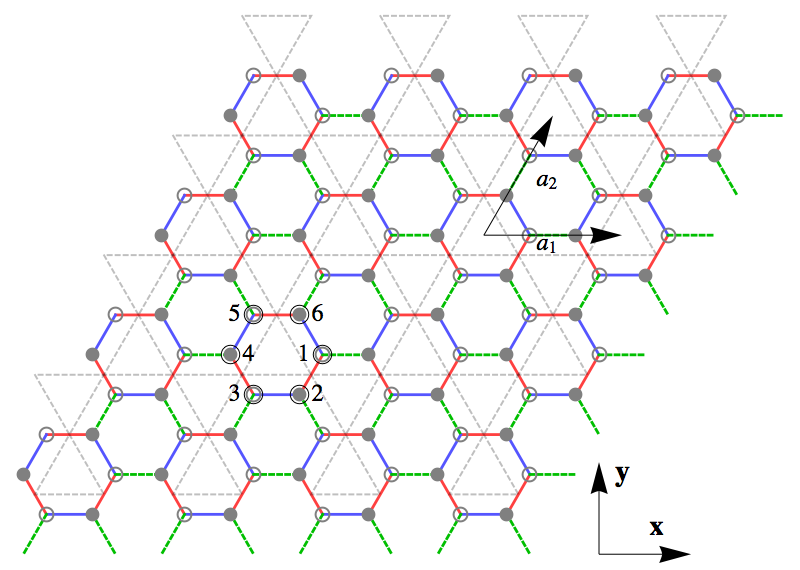}
\caption{(Color online) The Kekul\'{e}-Kitaev model.\cite{1742-5468-2010-08-P08010} The blue (dark), green (dashed) and red (light) links are the $x, y$ and $z$ links respectively. There are six sites in the unit-cell as shown (denoted by $i=1,2,3,4,5,6$). We call the sublattices $1,3,5 (2,4,6)$ as {\it odd(even)} sublattices. Connecting the mid-point of any particular set of links (say $z$ links) gives a Kagome lattice (shown in dotted fray lines). The lattice vectors are : ${\bf a}_1=\{3,0\};~~~{\bf a}_2=\left\{\frac{3}{2},\frac{3 \sqrt{3}}{2}\right\}$.}
\label{fig_rotated}
\end{figure}

To gain a comprehensive understanding of generic QSL phases, it is useful to understand the features of exactly solvable Hamiltonians which survive the presence of perturbations that spoil their exact solvability. Furthermore, when such perturbations are sufficiently strong they can give rise to quantum phase transitions  by destabilizing the QSL. Thus these systems present microscopic settings to study quantum phase transitions out of a QSL phase, an area which is far from well understood. From the material perspective, systematic study of the influence of such perturbations, which are inevitably present in candidate material systems, is also an imperative issue. Motivated by the above questions, in this paper we study an example  of a concrete spin Hamiltonian that exhibits an exactly solvable QSL  ground state and additional interactions lead to a continuous quantum phase transition to a magnetically ordered state. We systematically study the nature of this continuous quantum phase transition out of the QSL.

The spin Hamiltonian that we study is a variant of the exactly solvable Kitaev model (Fig. \ref{fig_rotated}) on a honeycomb lattice, and we analyze the effect of introducing additional antiferromagnetic Heisenberg interactions and also a magnetic field.  In the present model the distribution of the $x,y$ and $z$ type bonds form a Kekul\'{e} pattern as shown in Fig. \ref{fig_rotated}. This leads to important differences from the original construction of Kitaev, with interesting consequences in the structure of the phase diagram arising already in the exactly solvable limit. In particular, the model reduces to a toric code model on a Kagome lattice in an appropriate anisotropic limit. Using a combination of analytical and numerical approaches, we show that the effect of the Heisenberg interactions are quite different in the present case from the by now well known usual Heisenberg-Kitaev model.\cite{PhysRevLett.105.027204} In particular, we describe a magnetically ordered phase stabilized by a quantum `order by disorder' mechanism\cite{refId0} and a continuous quantum phase transition between this ordered phase and a $\mathbb{Z}_2$ QSL in the toric code limit. We construct the field theory which suggests that the critical point belongs to the $3D$-$XY\times Z_2$ universality class and support this by Monte Carlo simulations. Thus this model presents a controlled microscopic setting for a continuous quantum phase transition between a phase with collinear magnetic order and a $\mathbb{Z}_2$ QSL, in itself a subject of much recent interest.\cite{PhysRevB.86.214414,PhysRevB.89.235122}

The rest of the paper is organized as follows. We start with a brief introduction to the Kekul\'{e}-Kitaev model in Sec. \ref{sec_honey} and outline the basic features of the exact solution, the phase diagram, and point out the important differences with the original Kitaev model. We then discuss an interesting limit of the model, the so called ``strong-bond limit'', which leads to the a toric code model on a Kagome lattice. In Sec. \ref{sec_hkmodel}, we  add an antiferromagnetic Heisenberg interaction to the above Kitaev model and investigate the  stability of the QSL. While the QSL is stable to the weak short ranged spin-spin interactions, as expected, they lead to interesting phase transitions when they become sufficiently strong. In particular we find that in the toric code limit such interactions lead to a magnetic ordering that is stabilized by a quantum `order by disorder' mechanism. The transition between the QSL and the magnetically ordered phase is continuous. Using a combination of field theoretic arguments and Monte-Carlo calculations we find that this continuous transition belongs to the $3D$-$XY\times Z_2$ universality class. The effect of an external magnetic field is studied in Sec. \ref{sec_zeeman} where we find that unlike the usual Kitaev model, the present one does not harbour a chiral spin liquid at small magnetic field. We summarize our results in Sec. \ref{sec_discuss}. Calculational details are discussed in the appendices.
\section{The Model}
\label{sec_honey}

We start by outlining the spin-1/2 Kekul\'{e}-Kitaev model on the honeycomb lattice. Kitaev, in his pioneering work,\cite{kitaev2006anyons} considered a spin model on a honeycomb lattice where, depending on the direction of the three nearest neighbours, there are three types of spin exchanges.  As pointed out by Kamfor {\it et. al.},\cite{1742-5468-2010-08-P08010} Kitaev's original construction of the exactly solvable model can be extended to other types of distributions of the bond types on the honeycomb lattice. The general Kitaev Hamiltonian is given by
\be
\mathcal{H}_{\rm K}=-\sum_{\langle ij\rangle-\rm \a~ links}J_\a\s_i^\a\s_j^\a 
\label{eq_kitaev_r}
\ee
where $\s^\a_i~(\a=x,y,z)$ are the Pauli matrices representing the spin-1/2 at the site $i$, and the summation runs over the links of the honeycomb lattice, which are of three types ($\alpha=x,y,z$). Here, following Kamfor {\it et. al.},\cite{1742-5468-2010-08-P08010} we consider a different distribution of the three types of bonds compared to Kitaev's original model.\cite{kitaev2006anyons} This is depicted in Fig. \ref{fig_rotated}. There are three distinct types of hexagonal plaquette, which we denote as: (1) $x$-plaquettes where the bonds alternate between $y$ and $z$ types, (2) $y$-plaquettes where the bonds alternate between $x$ and $z$ types, and (3) $z$-plaquettes where the bonds alternate between $x$ and $y$ types. The links of a given type are therefore not parallel, but instead form a Kekul\'e type of pattern, and so we refer to the model as the Kekul\'{e}-Kitaev model. 

The distribution of links requires the unit cell to contain six sites (see Fig.\ref{fig_rotated}). We choose the $y$-plaquette as the unit cell. These plaquettes form a triangular lattice. The Brillouin zone information, along with its connection to the Brillouin zone of the underlying honeycomb lattice, are given in Appendix \ref{appen_bz} (Fig. \ref{fig_bz_rel}). The symmetries of the above Hamiltonian (see Fig. \ref{fig_rotated}) include -- (1)  lattice translation along the lattice vectors: ${\bf a}_1, {\bf a}_2$, (2) $2\pi/3$ rotation about the plaquette centre, (3) reflection about a line connecting the bond centres that lie on the opposite side of the plaquette, and (4) time reversal. In addition, the model has an extra symmetry along the isotropic line $J_x=J_y=J_z$. This is composed of a simultaneous inversion of the lattice about an $\alpha$-bond and a global rotation of the spins by $\pi$ about the $\a$-axis ($\alpha=x,y,z$).
\paragraph*{The exact solution :}

The exact solution of the above model is analogous to that of the usual Kitaev model.\cite{kitaev2006anyons} Here we outline the essential features, and relegate further details to Appendix \ref{appen_lieb}. As in Kitaev's original construction we first identify conserved plaquette operators. Since the lattice of the present model contains three different types of plaquettes, we define three types of  plaquette operators
\begin{align}\la{plaq_ops}
\mathcal{W}_\alpha(P)=-\prod_{i\in P}\sigma^\alpha_i=\prod_{ij\in\beta-link,\in P}\sigma_i^\beta\sigma_j^\beta~~~~~~~~~(\beta\neq\alpha)\end{align}
with $\alpha=x,y,z$. 
These differ from those introduced in Ref. \onlinecite{1742-5468-2010-08-P08010}  by an overall minus sign. By construction, these plaquette operators commute with the Hamiltonian 
as well as among themselves. 
Hence the Hamiltonian has  an infinite set of conserved quantities which are the $Z_2$ fluxes through the plaquettes. 
These conserved fluxes give rise to flux sectors, each of which has dimension $2^{N_{site}/2}$. 
In Appendix \ref{appen_lieb}, we show that Lieb's theorem\cite{PhysRevLett.73.2158} can be used to establish that the ground state lies in the zero-flux sector, where $\mathcal{W}_p^\alpha=+1$ for all plaquettes.
The remaining details of the solution proceed exactly as in Kitaev's construction\cite{kitaev2006anyons} and are outlined  in Appendix \ref{appen_lieb}.

\begin{figure}
\centering
\includegraphics[scale=0.35]{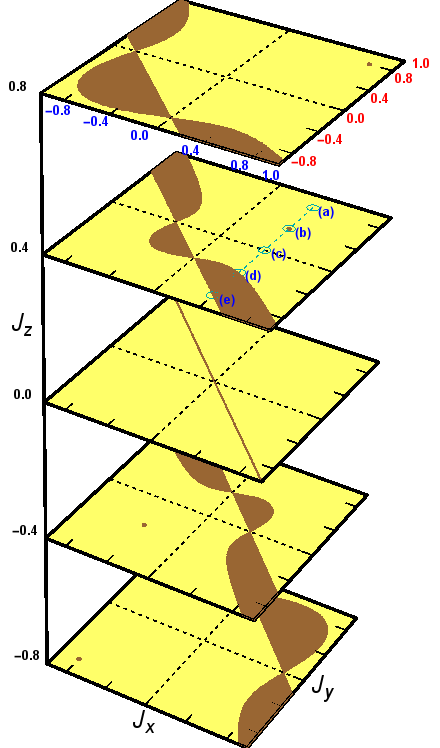}
\caption{(Color online) The phase diagram of the Kekul\'{e}-Kitaev model (Eq. \ref{eq_kitaev_r}) as a function of the coupling constants. The brown and yellow regions represent gapless and gapped phases respectively. For isotropic couplings ($J_x=J_y=J_z$), the system is always gapless, but it becomes gapped for infinitesimal perturbations away from this limit (see text for details). The band structures for the cut (a-e) are shown in Fig. \ref{fig_band}.}
\label{fig_pd}
\end{figure}

The structure of the phase diagram in the present case is however quite different from that of Kitaev's original construction, due to the difference between the symmetries of the two models. The phase diagram is shown in Fig. \ref{fig_pd}, where the brown  and yellow regions denote gapless and gapped phases respectively. The gapless region includes the plane given by the equation  $J_x+J_y+J_z=0$. In terms of the Majorana $c$ fermions (see Appendix \ref{appen_lieb}), representative dispersion curves are presented in Fig. \ref{fig_band}. A noteworthy feature is that the Majorana $c$ fermions are gapless along the isotropic line $J_x=J_y=J_z$, as seen in Fig. \ref{fig_pd}. Along this line, the Majorana $c$ fermions of the zero-flux sector have a nearest neighbour tight-binding Hamiltonian with a Dirac point occurring at the $\Gamma$ point of the folded Brillouin zone (refer to Fig. \ref{fig_bz_rel}), which is is protected by the special inversion symmetry present along this isotropic line.  Once we move even infinitesimally away from this isotropic line the Majorana fermions gain a  mass. The anisotropy so generated is similar to the Kekul\'e superconducting order parameter discussed in context of graphene.\cite{PhysRevB.82.035429} Details on the structure of the mass term for the low energy theory are given in Appendix \ref{appen_lieb}.
\subsection{The strong bond limit: toric code on the Kagome lattice}\la{ToricLimit}

An interesting limit of the present model is obtained when one of the couplings (say $J_z$) is much stronger than the other two. This is the so called toric code limit.\cite{kitaev2006anyons,PhysRevLett.90.016803} 
Here the $Z_2$ fluxes ($\mathcal W^\alpha_p$) provide the low energy degrees of freedom, as the $c$-Majorana fermions (see Appendix B) have a large gap ($\mathcal O(J_Z)$) in this limit.
\cite{kitaev2006anyons,PhysRevB.84.115146}
To obtain an effective description, we first consider the extreme limit $J_x=J_y=0$, and  $J_z>0$. The lattice separates into disjoint $z$-bonds with an Ising coupling term $-J_z\sigma_i^z\sigma^z_j$. Each bond has doubly degenerate ground states  $|\uparrow\uparrow\rangle$, $|\downarrow\downarrow\rangle$, as well as two high energy states 
$|\uparrow\downarrow\rangle$, $|\downarrow\uparrow\rangle$. 
We introduce the bond doublets\cite{kitaev2006anyons}
\begin{align}\la{bond_doublets}
|\Uparrow\rangle=|\uparrow\uparrow\rangle,~~~~~~|\Downarrow\rangle=|\downarrow\downarrow\rangle
\end{align}
to represent the ground state subspace on each bond. We adopt the notation that the first (second) spin always belongs to a site of the  sub-lattice of the honeycomb lattice that is denoted by open (solid) circles in Fig. \ref{fig_rotated}.  It is useful to note that under time-reversal symmetry: $|\Uparrow\rangle\rightarrow|\Downarrow\rangle$, $|\Downarrow\rangle\rightarrow|\Uparrow\rangle $. Introducing Pauli matrices $(\tau^x,\tau^y,\tau^z)$ on this bond-doublet space, time reversal is affected by $\mathcal{T}=\tau^x\mathcal{K}$, where $\mathcal{K}$ is the complex conjugation operator. This acts as $\mathcal{T}:~\{\tau^x,\tau^y,\tau^z\}\rightarrow\{\tau^x,\tau^y,-\tau^z\}$, and so $\tau$ represent a non-Kramers doublet.

The toric code appears once small $J_x,J_y$ couplings are taken into account, and can be obtained using degenerate perturbation theory on the bond doublets. This gives rise to an effective model on the Kagome lattice, which is formed by joining the midpoints of the $z$-bonds (refer to Fig. \ref{fig_rotated}). The details of the effective Hamiltonian are given in Appendix \ref{appen_toric_pert},\cite{1742-5468-2010-08-P08010,Kamfor} giving (up to constants, and at leading nonzero order, up to 6th order in perturbation theory):
\begin{align}
H_{TC} =& -\frac{{ 3} J_x^3}{8J_z^2}\sum_{\bigtriangleup} \lF+\frac{{ 3}  J_y^3}{8J_z^2}\sum_{\bigtriangledown} \rF -\frac{ 3 J_x^3 J_y^3}{256J_z^5} \sum_{\hexagon} \hF\nonumber\\
&+  \frac{ 7 J_x^3 J_y^3}{64 J_z^5}\sum_{\langle\bigtriangleup,\bigtriangledown \rangle } \lF \rF\,.
\label{eq_toric}
\end{align}
Here the sum $\langle,\rangle$ is  taken over corner-sharing  pairs of triangles, and we have introduced the notations 
\be
\lF = \prod_{K\in \bigtriangleup} \t^x_K\,,\quad \rF = \prod_{K\in \bigtriangledown} \t^x_K\,,\quad \hF = \prod_{K\in \hexagon} \t^z_K\,,
\ee 
where the subscript $K$ refers to the sites of the Kagome lattice. The plaquette operators $\lF$, $\rF$, $\hF$ all mutually commute and so the Hamiltonian can be diagonalized simultaneously with them. The eigenvalues of the plaquette operators, $\pm 1$, are good quantum numbers, and they specify the ground states as well as the excited states. We note  that  the triangle terms do not break time reversal symmetry, although they are of the form $\tau^x_I\tau^x_J\tau^x_K$, because $\tau$ is a non-Kramers doublet as discussed above. While the properties of the toric code model are known in context of the square lattice, we shall briefly summarize the details here for the sake of continuity.

The Kagome lattice has a unit cell comprised of three sites (for example the up-triangle in Fig. \ref{fig_rotated}), and lattice vectors ${\bf a}_1$, ${\bf a}_2$. If $N_1$ and $N_2$ are the linear dimensions along the two crystalline directions, then the total number of up-triangles is $N_T=N_1N_2$. There are $3N_T$ Kagome lattice sites, and thus  the Hilbert space spanned by the $\tau$ spins has a dimension of $2^{3N_T}$. Correspondingly there are $N_T$ of each of the $\lF$, $\rF$ and $\hF$ operators, and  specifying their values also leads to $2^{3N_T}$ states. However, the operators are not all independent on a 2-tori, as they obey two constraints:
\be\la{2tori_constr}
\Big(\prod_{\bigtriangledown}\rF\Big) \Big(\prod_{\bigtriangleup}\lF\Big)=+1\,, \quad \prod_{\hexagon}\hF=+1\,.
\ee
These give rise to the topological degeneracy of $4$, as  expected for a gapped $Z_2$ quantum spin liquid in a toric code model.\cite{kitaev2003fault}

On taking $J_y=0$, the lattice splits into disconnected up-triangles. There are eight states per triangle with a  four-fold ground state degeneracy. The ground state degeneracy for  a lattice of $3N_T$ sites is then $4^{N_T}=2^{2N_T}$. This extra degeneracy is accidental and is immediately lifted when $J_y$ is turned on. For $J_y<0$ (we shall always take $J_x>0$), the ground state lies in a sector where the eigenvalues of each of the plaquette operators are $\lF=\rF=-\hF=+1$, as this minimises each of the four terms in the Hamiltonian. For even $N_T$, the ground state has the  $4$-fold topological degeneracy referred to above. For odd  $N_T$ however, the constraints of Eq. \eqref{2tori_constr} do not allow all hexagons to have $\hF=-1$, and so the ground state must have one of them to be +1. This ``defect" honeycomb can sit anywhere on the 2-tori and the ground state degeneracy is raised to $4N_T$. A similar feature is also observed for $J_y>0$.

\begin{figure}
\centering
\subfigure[]{
\includegraphics[scale=0.175]{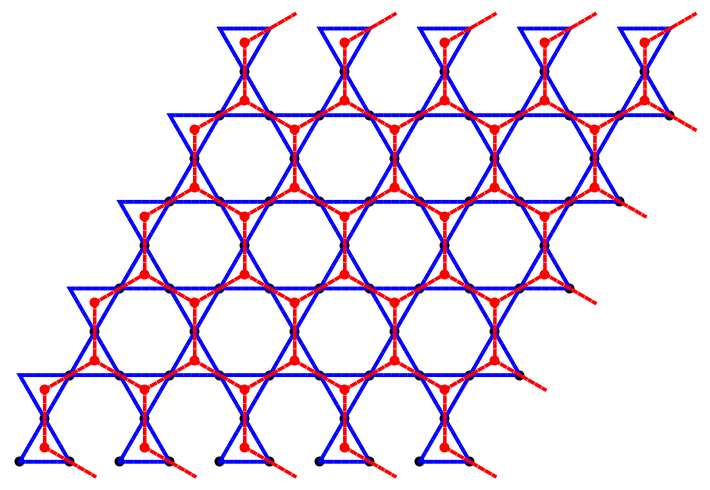}
\label{fig_medial}
}
\subfigure[]{
\includegraphics[scale=0.175]{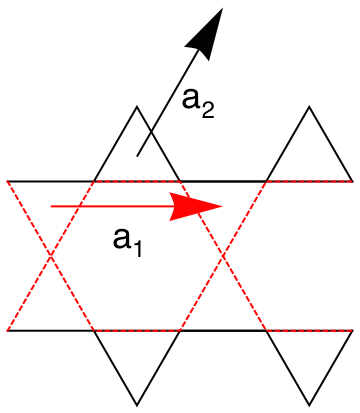}
\label{fig_boundary}
}
\caption{(Color online) (a) The medial honeycomb lattice (red) of the Kagome  lattice (blue). (b) A boundary can be obtained by setting the coupling strengths along the red (dashed) triangles and honeycombs to zero.}
\end{figure}

In the bulk there are two types of excitations, both of which are gapped and dispersionless.\cite{kitaev2006anyons} There are Ising electric charges which are associated with the triangular plaquettes, and Ising magnetic charges associated with the hexagonal plaquettes. To study these excitations it is instructive to go to the medial lattice of Kagome, which is obtained by joining the centres of neighbouring triangular plaquettes as shown in Fig. \ref{fig_medial}. This gives another  honeycomb lattice, which is different to  that of Fig. \ref{fig_rotated}. The excitations can now be understood as the endpoints of string operators. A pair of electric charges are created by the operator
\begin{align}
W^{(e)}_{\ell }=\prod_{I\in \ell}\tau^z_I,
\end{align}
where $\ell$ denotes a path on the honeycomb lattice starting and ending at the two charges, while a pair of magnetic charges are created by
\begin{align}
W^{(m)}_\ell=\prod_{I\in \ell}\tau^x_I,
\end{align}
where here $\ell$ is a path on the triangular lattice obtained by connecting the centres of the honeycomb plaquettes. Examining how the charges wind around one another,  it is  straightforward to see that the electric and magnetic charges are bosons with mutual semionic statistics, i.e. each sees the other as source of $\pi$ flux.\cite{kitaev2006anyons} We note that while the electric charges move on a bipartite (honeycomb) lattice, the magnetic charges move on the non-bipartite (triangular) lattice. 

\paragraph*{Open boundary conditions and Majorana edge modes:}

The system has gapless edge states in the presence of an open boundary, which are related to the underlying topological order.\cite{PhysRevLett.90.016803} To describe these, let us imagine drawing a great circle around the system on a 2-tori, obtained by translation along ${\bf a}_1$  as shown in Fig. \ref{fig_boundary}, and then cutting the system by setting all the $\lF, \rF$ and $\hF$ crossed by the great circle to be zero. Let us again take $N_1$ and $N_2$ as the linear dimensions, so there are $3N_1N_2-N_1$ connected Kagome lattice sites, and the numbers of each of up and down triangles and hexagons are $N_1(N_2-1)$, leading to a degeneracy of  $2^{3N_1N_2-N_1}/2^{3N_1(N_2-1)}=2^{2N_1}$. On the other hand there are $4N_1$ edge spins, $2N_1$ on each edge. This gives rise to $\sqrt{2}$ degree of freedom at each edge site, which corresponds to Majorana edge states. These edge states have a flat band with exactly zero energy and are stable to weak perturbations away from the exactly solvable limit as long as translational symmetry along the edge is not spontaneously broken.
There are other more complicated edges that could be considered, but this is outside the scope of the present work.

\paragraph*{Distinct toric codes and adiabatic continuity:}

In the above treatment we have focused on the toric code model that appears in the  $J_z\gg J_x,J_y$ regime. However we could have done the same for the other two couplings, $J_x$ and $J_y$. In Kitaev's original model\cite{kitaev2006anyons} the three toric codes so obtained cannot be connected without closing the bulk excitation gap. However, we find that in the present model there exist only two such distinct toric code limits. 
These are separated by the $J_x+J_y+J_z=0$ plane and are adiabatically connected to the two toric codes obtained in the vicinity of the two limits $J_z\rightarrow \pm \infty$. 

To demonstrate this, imagine setting $J_x=J_y=0$ and $J_z=1$, which breaks the honeycomb lattice of Fig. \ref{fig_rotated} into disjoined bonds (which are the sites of the Kagome lattice). Turning on small $J_x$ only results in disjoined honeycombs (or isolated up-triangles for the Kagome lattice). This is contrary to the original Kitaev model where turning on two bonds results in extended 1D chains. Thus in the present case, even for finite $J_x$, there is no dispersion when $J_y=0$. For the disjoined honeycombs, albeit with anisotropic exchanges on their bonds, the six eigenstates have energies $\pm (J_x + J_z)\,, \quad \pm \sqrt{J_x^2 - J_x J_z + J_z^2}\,, \quad \pm \sqrt{J_x^2 - J_x J_z + J_z^2}\,.$ For any value of $J_x>0$ the gap never closes. The wavefunctions however become superpositions of the $\sigma$ spin states on the six sites of the hexagon containing $x$ and $z$ bonds. On attaining the point $J_x=J_z=1$, one can then take $J_z$ to zero gradually without closing the gap, resulting again in disjoined bonds, but now the doublets are polarized parallel to the $x$ direction.
So it is possible to continuously connect the two bond limits without closing the bulk energy gap, and hence to adiabatically connect the two toric codes which arise from perturbations about these two limits. 
This argument does not work however if we try and connect  $J_x=J_y=0$, $J_z=1$ to  $J_z=J_y=0$, $J_x=-1$, as then a gap closing point is necessarily encountered.
It follows then that the two toric code phases obtained in the vicinity of $J_z>0$  and $J_z<0$  cannot be continuously connected as the bulk gap closing and reopening necessarily occurs across the $J_x+J_y+J_z=0$ plane.

This completes our discussion of the phase diagram of the Kekul\'{e}-Kitaev model. Starting from the next section we shall investigate the effect of perturbations.

\section{Heisenberg perturbations}
\label{sec_hkmodel}

An important class of interactions in effective spin models are bilinear spin-spin interactions. To this end we now investigate the effect of adding antiferromagnetic Heisenberg interactions to the Kekul\'e-Kitaev model, and shall see that this leads to interesting consequences. 
We thus turn our attention to the following extended Hamiltonian
\begin{align}
H=J_H\sum_{\langle ij\rangle}\mathbf{\sigma}_i\cdot\mathbf{\sigma}_j+ \mathcal{H}_{\rm K}\,,
\label{eq_hk}
\end{align}
where $\mathcal{H}_{\rm K}$ is given by Eq. \ref{eq_kitaev_r}. 
As we shall see, the resulting phase diagram is quite rich and allows for interesting phases and phase transitions, the most interesting of which is a continuous transition between the QSL and a magnetically ordered phase. In this paper, we shall concentrate on the toric code limit for the Kitaev interactions and obtain a controlled description of not only the magnetic ordering driven by the Heisenberg perturbation, but also of a continuous quantum phase transition between the QSL and the magnetically ordered phase.

\subsection{The generalized toric code model}

Starting with Eq. \ref{eq_hk}, in the limit $J_H,J_x,J_y\ll J_z$, we obtain a generalized toric code model with nearest neighbour antiferromagnetic Ising perturbations. The hierarchy of energy scales allow for a strong coupling perturbative expansion, as in the previous section, yielding the following  effective Hamiltonian to the leading order in all the couplings
\begin{align}
H=&a\sum_{\langle IJ\rangle} \tau^z_I\tau^z_J-a_\bigtriangleup\sum_\bigtriangleup \mathcal{F}_\bigtriangleup-a_\bigtriangledown\sum_\bigtriangledown \mathcal{F}_\bigtriangledown-a_{\hexagon}\sum_{\hexagon} \mathcal{F}_{\hexagon}\nonumber\\
&-a_{\bigtriangleup\bigtriangledown}\sum_{\langle\bigtriangleup,\bigtriangledown\rangle}\mathcal{F}_\bigtriangleup\mathcal{F}_\bigtriangledown\,.
\label{eq_gen_toric}
\end{align}
This is a generalized toric code model where the last four terms are same as the toric code Hamiltonian (Eq. \ref{eq_toric}), albeit with renormalized couplings, and the first term is a nearest neighbour antiferromagnetic Ising interaction between the $\tau$ spins sitting on the Kagome lattice, and $a=J_H$ to leading order.

A special feature here is that $\tau^z$ remains a conserved quantity, which is not the case in the original version of the Kitaev model.\cite{PhysRevB.90.035113} As a result, the magnetic fluxes are still good quantum numbers. This will be very important for the rest of our analysis and will help reveal the nature of the phase transition in the present model, which has so far eluded analytic understanding in the original Kitaev-Heisenberg models even in the toric code limit.\cite{PhysRevB.90.035113} The perturbations do however cause  the electric charges to acquire dynamics. 

To see the effect of the Ising term more clearly, it is useful to turn once more to the medial lattice construction shown in Fig. \ref{fig_medial}. The resulting honeycomb lattice (not to be confused with the lattice of Fig. \ref{fig_rotated}) has two sublattices which are respectively at the centres of the up and the down triangular plaquettes of the Kagome lattice. The Ising term leads to hopping of electric charges on the honeycomb lattice preserving the sublattice flavour, i.e. next-nearest-neighbour hopping. Since the electric charges are bosons, they can condense once their dispersion minimum touches zero. We shall show that this phase breaks time reversal symmetry and hence generates magnetic order. 

Let us first generalise our viewpoint. While the Hamiltonian of Eq. \eqref{eq_gen_toric} has been derived from a microscopic model with a definite hierarchy of the coupling constants, we shall relax that hierarchy for the moment and consider the generalized phase diagram for the above model. We shall comment on the actual microscopic parameters at the end in context of the generalized phase diagram.

To proceed we introduce a gauge theory description of the generalized model, by defining the following set of Ising variables\cite{PhysRevLett.98.070602} on the medial honeycomb lattice (Fig. \ref{fig_medial})
\begin{align}
\lF=\mu^x_p,\quad \rF=\mu^x_q,\quad \tau_I^z=\mu^z_p\rho^z_{pq}\mu^z_q\,,
\label{eq_transform}
\end{align}
where $\mu$ spins are defined on the sites and $\rho^z$ are defined on the links. The $\mu^z=\pm1$ carry Ising gauge charge and $\rho^z=\pm1$ are the Ising gauge potentials.
Here $p,q$ are the nearest neighbour sites on the medial honeycomb lattice, and so belong to the two different sublattices of it.  The  Hamiltonian of Eq. \ref{eq_gen_toric} now takes the form
\begin{align}
H&=a\sum_{\langle\langle pp'\rangle\rangle,q}\mu^z_p\rho^z_{pq}\rho^z_{qp'}\mu^z_{p'}-a_\bigtriangleup\sum_p\mu^x_p-a_\bigtriangledown\sum_q\mu^x_q\nonumber\\
&~~-a_{\hexagon}\sum_{\hexagon}\prod_{pq\in\hexagon}\rho^z_{pq}-a_{\bigtriangleup\bigtriangledown}\sum_{\langle pq\rangle}\mu^x_p\mu^x_q\,,
\label{eq_gen_toric_gauge}
\end{align}
where in the first term $q$ is the common nearest neighbour site connecting the two  second nearest neighbour sites $p,p'$.
The Ising magnetic fluxes are  given by
\begin{align}
\prod_{pq\in\hexagon}\rho^z_{pq}=-1\,,
\end{align}
 and they sit on a triangular lattice formed from the plaquettes of the medial honeycomb lattice.

In the limit $a_\bigtriangleup,a_\bigtriangledown,a_{\bigtriangleup\bigtriangledown}=0$ and $a, a_{\hexagon}>0$, the Hamiltonian becomes classical as it has only $\tau_z$ operators.
The most degenerate point of parameter space is obtained on further setting  $a_{\hexagon}=0$. This corresponds to  a classical antiferromagnetic Ising model on the Kagome lattice, and has a ground state entropy of $0.502~k_B$ per site of the Kagome lattice.\cite{kano1953antiferromagnetism,liebmann1986statistical,pauling1960nature} The degeneracy is partially lifted however by  an infinitesimally positive $a_{\hexagon}$. This chooses the ground state sector which has zero magnetic flux through the hexagonal plaquettes. This is easiest to see in the gauge theory, where $a_{\hexagon}$ forces the magnetic flux through the hexagons to be zero. As a result, a gauge with all $\rho^z=+1$ can be chosen, yielding two copies of the triangular lattice antiferromagnet, each of which has an entropy of $0.323~k_B$ per site of the triangular lattice.\cite{kano1953antiferromagnetism,liebmann1986statistical,pauling1960nature} As there are three Kagome lattice sites for each triangular lattice site, the resulting entropy is $0.215~k_B$ per site of the Kagome lattice. The quenching of the entropy comes from the fact that $\tau^z$ spin configurations whose product around the hexagons is $-1$ are now energetically more costly and hence not in the ground state manifold. The entropy nevertheless remains macroscopic due to the absence of quantum fluctuations, to which we now turn our attention.

\begin{figure*}
\centering
\subfigure[]{
\includegraphics[width=0.8\columnwidth]{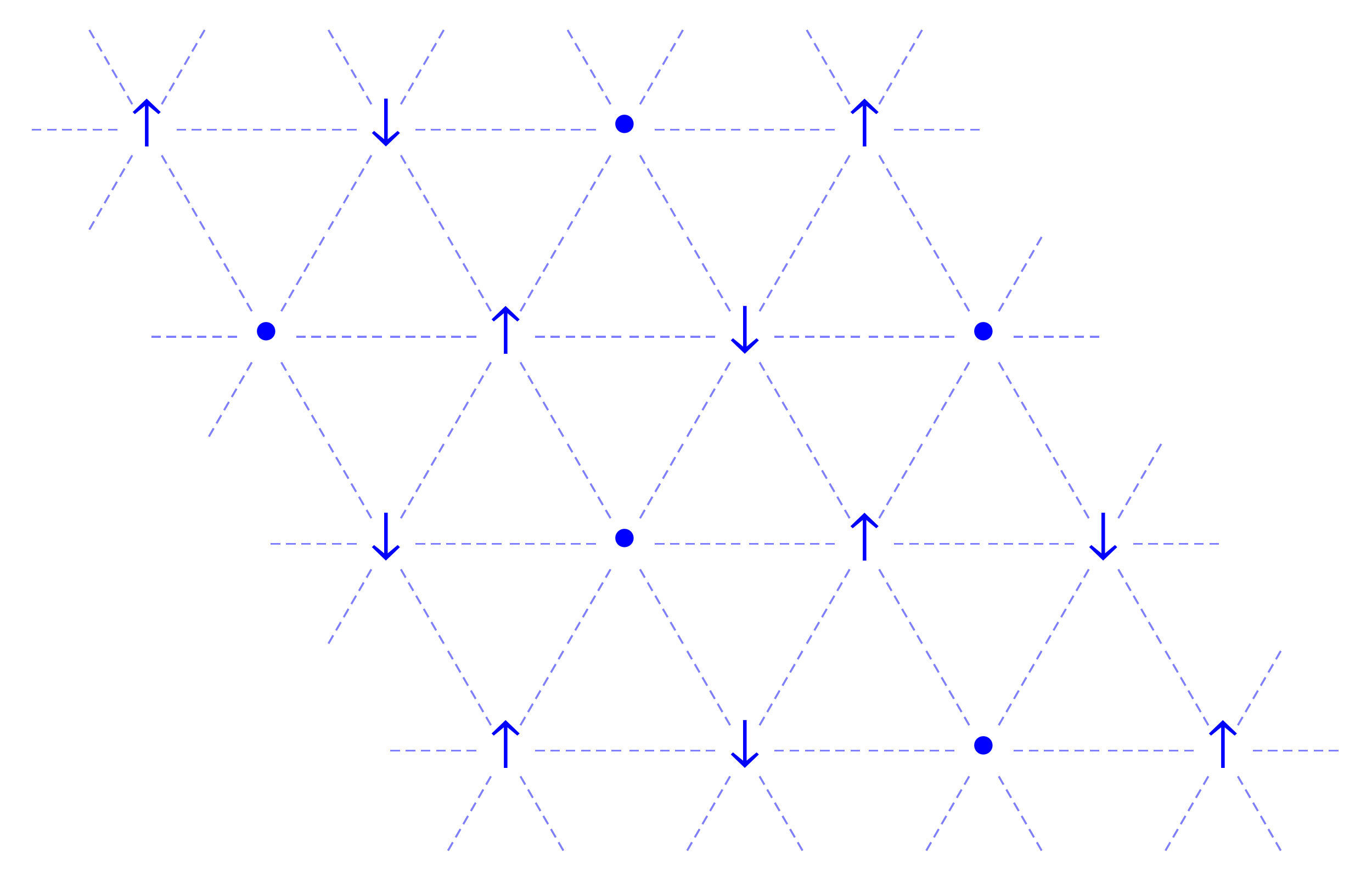}
\label{fig:tri_order}
}
\subfigure[]{
\includegraphics[width=0.8\columnwidth]{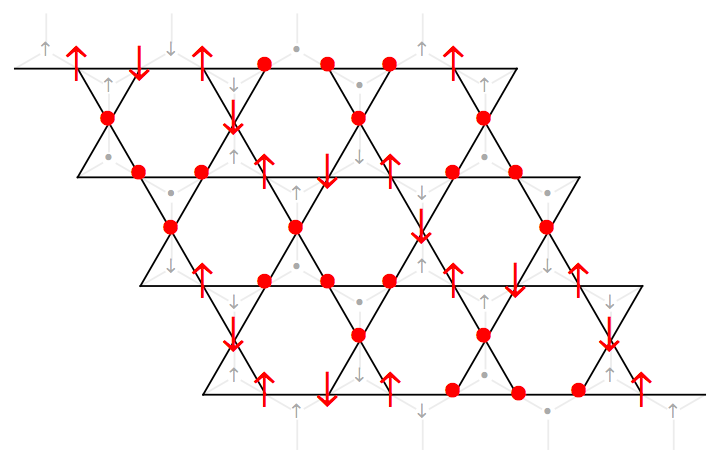}
\label{fig_order_kag}
}
\caption{(Color online) (a) Magnetic ordering for the antiferromagnetic TFIM on the triangular lattice. The dots refer to sites which have zero magnetization (the maximally flippable sites)
 (b) Magnetic ordering of the $\tau^z$ spins on the Kagome lattice.} 

\end{figure*}

To consider the quantum fluctuations, let us start with the limit $a_\bigtriangleup,a_\bigtriangledown,a_{\bigtriangleup\bigtriangledown},a\ll a_{\hexagon}$, with all being positive. In this limit we can concentrate on the sector where there are no magnetic charges, i.e. $\mathcal{F}_{\hexagon}=+1~\forall {\hexagon}$. Hence we can choose a gauge where $\rho^z=1$ for  all the links of the medial lattice. The Hamiltonian in Eq. \eqref{eq_gen_toric_gauge} can then be cast into a transverse field Ising model (TFIM)
\begin{align}
H_{TFIM}=&a\sum_{\langle\langle pp'\rangle\rangle}\mu_{p}^z\mu_{p'}^z-a_\bigtriangleup\sum_p\mu^x_{p}-a_\bigtriangledown\sum_q\mu^x_{q}\nonumber\\
&-a_{\bigtriangleup\bigtriangledown}\sum_{\langle pq\rangle}\mu^x_p\mu^x_q\,.
\label{eq_oneham}
\end{align}
In terms of the new variables, the spin liquid of Sec. \ref{ToricLimit} becomes the paramagnetic phase of the $\mu$ spins in which all the spins are polarized along $\mu^x$. When $a=a_{\bigtriangleup\bigtriangledown}=0$ this is clearly the ground state, and the basic excitations, the $Z_2$ electric charges (the flipped $\mu^x$ spins), are gapped with an energy cost of the order $a_\bigtriangleup$ or $a_\bigtriangledown$. When $a$ is turned on these $Z_2$ charges acquire dispersion, with a  bandwidth proportional to $a$. Once this becomes comparable to the gap, these charges condense at a particular momentum. In such a condensed phase $\langle\mu^z\rangle\neq 0$, and consequently $\langle \tau^z\rangle\neq 0$. Since $\tau^z$ is odd under time reversal, this phase breaks time reversal symmetry (along with lattice symmetries) and is actually a magnetically ordered state.

\paragraph*{The magnetic order:} Let us first describe this magnetic ordering for $a_{\bigtriangleup\bigtriangledown}=0$, as this decouples the Hamiltonian to two copies of the TFIM on the triangular lattice, a model which is well known.\cite{PhysRevB.29.5250,PhysRevB.68.104409,PhysRevB.63.224401,PhysRevLett.84.4457} 
In the classical limit this model has an extensive ground state degeneracy, albeit with power-law spin correlations.\cite{PhysRevB.63.224401} When the transverse field is turned on, the system goes into a magnetically ordered state through a quantum `order by disorder' mechanism.\cite{PhysRevB.45.7536,refId0} The ground state has a characteristic $\sqrt{3}\times\sqrt{3}$ order where there is a hexagonal lattice superimposed on the triangular lattice on which the spins exhibit Neel order in $\mu^z$ direction, while the spins at the site in the centre of each hexagon are polarized along the $\mu^x$ direction, and gain their energy from the transverse field by being in the maximally flippable state.  For each triangle  the three sites have magnetization, $M_z=\langle \mu^z\rangle$, of the form $(+1,-1,0)$. This candidate magnetic order in terms of the $\mu^z$ spins for a single triangular lattice is shown in Fig. \ref{fig:tri_order}. The ordering in terms of the $\tau^z$ spins can be obtained through Eq. \ref{eq_transform} and this is shown in Fig. \ref{fig_order_kag}. A characteristic feature of the ordering is the regular pattern of bow-ties with zero magnetization, which are surrounded by zig-zag chains of antiferromagnetic order. It is interesting to note that the chains of antiferromagnetic order are mutually decoupled from each other by the zero magnetization bow ties. In our Monte Carlo studies on the Hamiltonian \ref{eq_oneham} (described below) on a space-time lattice, we find the above ordering pattern persists throughout the part of the phase diagram that is  magnetically ordered even when the triangular lattices are coupled ($a_{\bigtriangleup\bigtriangledown}\neq0$).

Now we describe the transition between the QSL and this magnetically ordered state, which as we remarked earlier should be looked upon as a transition arising from the condensation of the Ising electric charge excitations of the QSL. Considering again one of the triangular sublattices, 
it is well known\cite{PhysRevB.29.5250} that the transition of the TFIM  can be effectively described by adopting a soft-spin description for the $\mu^z$ spins, and identifying the soft modes that condense to give rise to the magnetic order.
We follow a similar prescription for describing the present phase transition. The soft modes occur at ${\bf K}_\pm=\pm[4\pi/3,0]$, and so the spins expand as
\begin{align}
\mu^z({\bf r})=\psi_{+}({\bf r})e^{i{\bf K_+\cdot r}}+\psi_{-}({\bf r})e^{i{\bf K_-\cdot r}}\,,
\label{eq_soft_amp}
\end{align}
with amplitudes $(\psi_+,\psi_-)$. Taking all the global symmetries of the microscopic model into account, the $(2+1)$-dimensional Euclidean Landau-Ginzburg action can be constructed in the standard way (details are relegated to Appendix \ref{appen_ising}) giving\cite{PhysRevB.29.5250}
\begin{align}
\mathcal{S}_0[\vec{\psi}]=&\int d^3{\bf r}~\left[ |\partial\vec\psi|^2 + r_2 \vec\psi\cdot\vec\psi
+u_4(\vec\psi\cdot\vec\psi)^2+u_6(\vec\psi\cdot\vec\psi)^3\right.\nonumber\\
&\quad\quad\quad\left.+v_6 (\psi_+^6+\psi_-^6)\right]\,,
\label{eq_zeroLGW}
\end{align}
where the integration is over the $(2+1)$-dimensional Euclidean space-time, and
\begin{align}
\vec \psi=\{\psi_x,\psi_y\}=\left\{\frac{\psi_++\psi_-}{2},\frac{\psi_+-\psi_-}{2i}\right\}\,.
\end{align}
The above action predicts\cite{PhysRevB.29.5250} that the critical point belongs to the $3D$-$XY$ universality class, with the six-fold anisotropy term being dangerously irrelevant at the critical point. The sign of this anisotropy term determines the nature of the magnetic order. 

For $a_{\bigtriangleup}=a_\bigtriangledown$, the Hamiltonian has an inversion symmetry that exchanges the two sublattices of  the medial honeycomb lattice. In the rest of our calculations, we shall restrict ourselves to this inversion symmetric case and use $a_{\bigtriangleup}=a_\bigtriangledown=\Gamma$.  The limit $a_{\bigtriangleup\bigtriangledown}=0$ corresponds to two such decoupled triangular lattices. For each copy we introduce a pair of soft modes given by ${\vec \psi}^{(n)}$, where $n=1,2$ denotes the two copies, and so the action is
\begin{align}
\mathcal{S}_{\rm decoupled}=\sum_{n=1}^2\mathcal{S}_0[\vec{\psi}^{(n)}]\,.
\end{align}
The six-fold anisotropy term is again dangerously irrelevant at the critical point, and the symmetry is $3D$-$XY\times3D$-$XY$. In this limit the magnetic orders of the two triangular lattices are mutually independent. Symmetry however allows the coupling of the two triangular lattices and this is exemplified by the presence of the term $a_{\bigtriangleup\bigtriangledown}\neq 0$ in the generalized toric code model. We can write down the leading symmetry allowed term  (see again Appendix \ref{appen_ising} for details) to get the complete action
\begin{align}
\mathcal{S}=\mathcal{S}_{\rm decoupled}+\mathcal{S}_{\rm int}\,,
\label{eq_LGW}
\end{align}
where to the leading order
\begin{align}
\mathcal{S}_{\rm int}&=\int d^3{\bf r}\left[(\vec\psi^{(1)}\cdot\vec\psi^{(2)})\Big(r'_2+v_4\sum_{n=1,2}\vec\psi^{(n)}\cdot\vec\psi^{(n)}\right.\nonumber\\
&\quad\quad\quad\quad\quad\quad\left.+w_4(\vec\psi^{(1)}\cdot\vec\psi^{(2)})\Big)+v_4'|\vec\psi^{(1)}|^2|\vec\psi^{(2)}|^2\right]\,.
\end{align}
The term $\mathcal{S}_{\rm int}$ breaks the $3D$-$XY\times3D$-$XY$ symmetry down to  $3D$-$XY\times Z_2$ with a six-fold anisotropy term. The $Z_2$ symmetry is related to the fact that under inversion about the bond centre, the flavours of the soft-mode change. Tracing back to our microscopic Hamiltonian, this symmetry is present when $a_{\bigtriangleup}=a_\bigtriangledown$. Power counting, at the free fixed point, shows $r'_2,v_4,v_4'$ and $w_4$ are relevant while the sixth order terms ($u_6$ and $v_6$) are marginal. To understand their effect for the ordered phase ($r_2<0$) we look at the phase fluctuations. To this end we write
\begin{align}
{\vec \psi}^{(i)}=\rho(\cos\theta_i,\sin\theta_i)\,,
\end{align}
where we have taken the magnitude of the XY order parameter to be constant. Then, neglecting the six-fold anisotropy term, for the ordered phase we have\cite{PhysRevB.61.3430}  
\begin{align}
\mathcal{S}_{\rm decoupled}'=\int d{\bf r}^3\frac{K}{2}\left[|\partial\theta_+|^2+|\partial\theta_-|^2\right]\,,
\end{align}
where $\theta_\pm=(\theta_1\pm\theta_2)/2$, and
\begin{align}
\mathcal{S}_{\rm int}=\int d^3{\bf r}\left[r_2''\cos2\theta_- + \frac{w_4}{2}\cos4\theta_-\right]\,.
\end{align}
Now, depending on the sign of $r''_2$ and $w_4$, an expectation value for $\theta_-$ is chosen. Under inversion $\theta_-\rightarrow-\theta_-$, and so a transition which induces an ordering in $\theta_-$ simultaneously breaks the $Z_2$ symmetry related to inversion. Expanding the cosine terms around the $\theta_-$ expectation value makes $\theta_-$ excitations massive while $\theta_+$ remain massless. The latter is in fact the low energy XY degree of freedom.  Now the six fold anisotropy term becomes
\begin{align}
&v_6'\left[\cos[6(\theta_++\theta_-)]+\cos[6(\theta_+-\theta_-)]\right]\nonumber\\
&=2v_6'\cos(6\theta_+)\cos(6\theta_-)\,.
\end{align}
Replacing $\theta_-$ with its expectation value, we find that this term behaves like the six fold anisotropy field for $\theta_+$, which is known to be relevant in the low temperature ordered phase\cite{PhysRevB.61.3430,PhysRevB.16.1217,PhysRevB.29.5250}. 
 Hence we expect that the six-fold anisotropy is dangerously irrelevant at this transition, and that the critical point belongs to the $3D$-$XY\times Z_2$ universality class. This expectation is supported by our numerical studies below. Thus the field theory predicts that the transition due to the condensation of the Ising electric charges in the QSL belongs to an interesting universality class, and that the present microscopic model can harbour such an unconventional phase transition from a topological QSL phase to a magnetically ordered phase.

\paragraph*{Numerical calculations:}
To complement the above prediction of the soft mode analysis, we investigate the lattice model numerically. In particular we construct a discrete-time classical action for the spin model of Eq. \eqref{eq_oneham}, on which we perform Monte Carlo simulations.  The details of the derivation are given in Appendix \ref{appen_class}, and the resulting three dimensional Euclidean Landau-Ginzburg action, $E$, describes an Ising lattice model on a stacked honeycomb lattice with additional Ising link fields. This has the form:
 \bea\la{Eclmod}
J E=&J\sum_{\langle\langle ij\rangle\rangle,\tau} \mu^z_{i\tau}\mu^z_{j\tau}-\tilde{K}\sum_{i,\tau}\left(\prod_j\eta_{ij}\right)\mu^z_{i,\tau}\mu^z_{i,\tau+1}\\
&-K_\tau\sum_{\langle ij\rangle,\tau}\eta_{ij,\tau}\,.
\eea
where $\mu^z_{i}(=\pm 1)$ and $\eta_{ij}(=\pm 1)$ are Ising variables sitting on the sites and links of the stacked honeycomb lattice  respectively.  The couplings are related to those of the microscopic model as follows:
 $J=a\Delta\tau$, $\tanh(\tilde{K})=e^{-2\Delta\tau \Gamma}$, $\tanh(\Delta\tau a_{\bigtriangleup\bigtriangledown})=e^{-2K_\tau}$ (thus $K_\tau$ represents the coupling between the two triangular sublattices), and $\Delta\tau = \beta/N$, where $\beta$ is the inverse temperature (we are interested in the $\beta\to\infty$ limit) and $N$ is the number of temporal slices. Without loss of generality in the scaling limit, we study the isotropic case $\tilde K =J$ for the above model and explore the phase diagram as a function of the two dimensionless parameters ${1/J}$ and $K_\t/J$. The coupling $ J$ controls the strength  of the Heisenberg perturbations, while $K_\t$ controls the coupling of the two copies of TFIM. In addition, $1/J$ plays the role of temperature for the classical action. 

\begin{figure}
\centering
\includegraphics[width=\columnwidth]{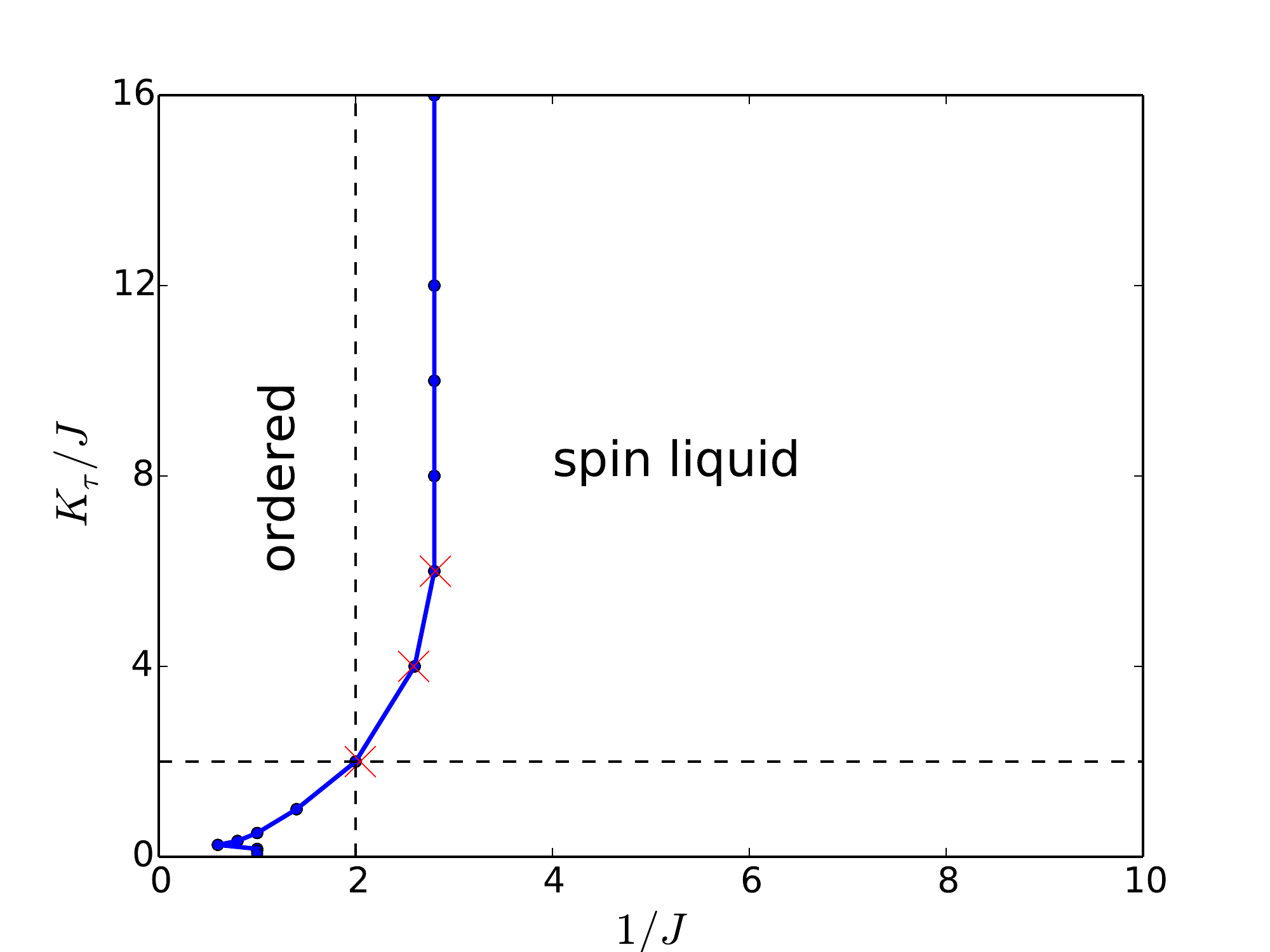}
\caption{(Color online) The phase diagram of the model \eqref{Eclmod} in the $1/J$-$K_{\tau}/J$ plane. 
The phase boundary is obtained from the peaks of the specific heat (blue dots), see Fig. \ref{fig:cT}. To ensure that these peaks correspond to the phase transition, critical couplings extracted from crossings of the Binder ratio are plotted (red crosses) for $K_{\tau}/J=2,4,6$, see Fig. \ref{fig_binder} for $K_{\tau}/J=2$.  For small $K_{\tau}/J$ the peak in the specific heat becomes rounded and the accuracy of the phase boundary is diminished. The two dashed lines represent the curves along which the magnetic correlations are plotted in Fig. \ref{fig:mag}. 
} 
\label{fig:cpd}
\end{figure}

\begin{figure}
\centering
\subfigure[]{
\includegraphics[width=\columnwidth]{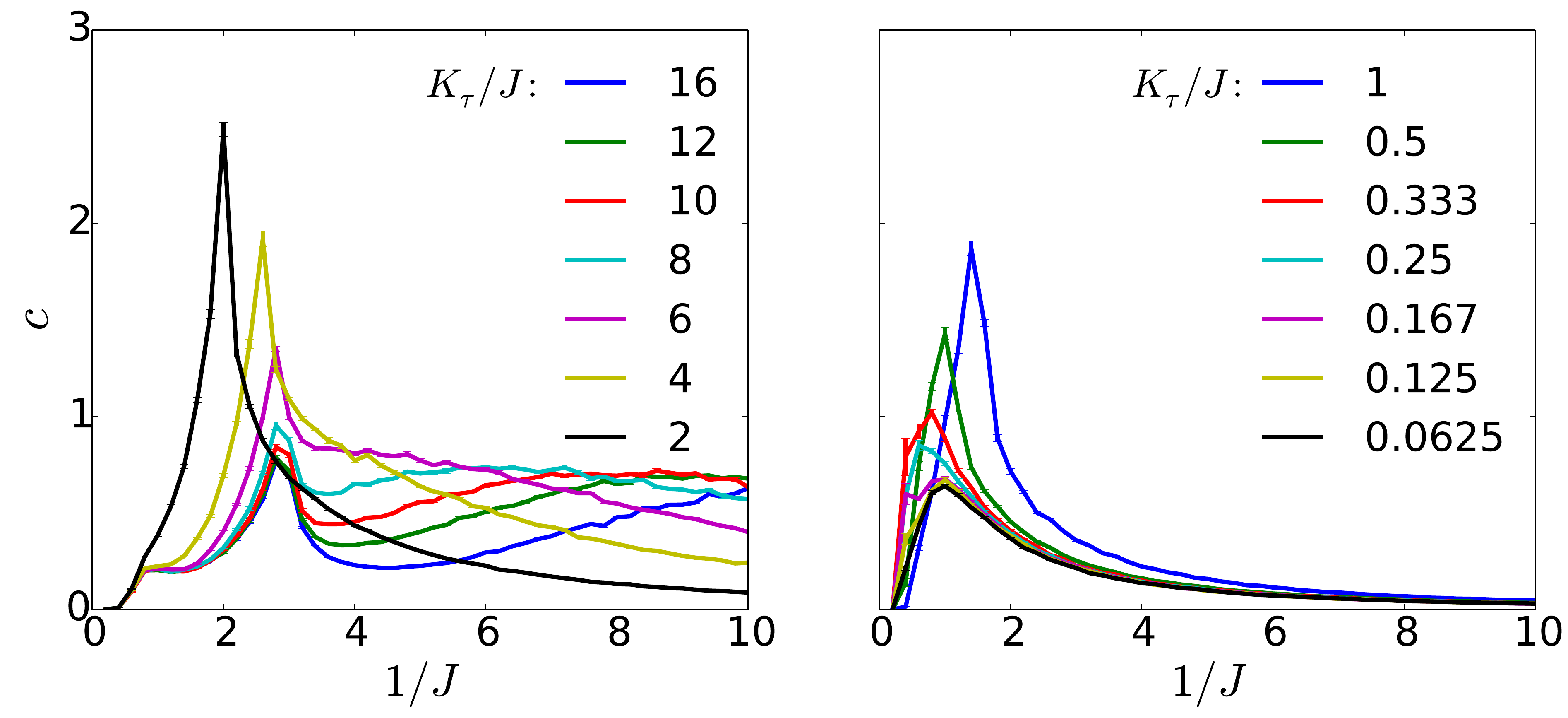} 
\label{fig:cT}
}
\subfigure[]{
\includegraphics[width=\columnwidth]{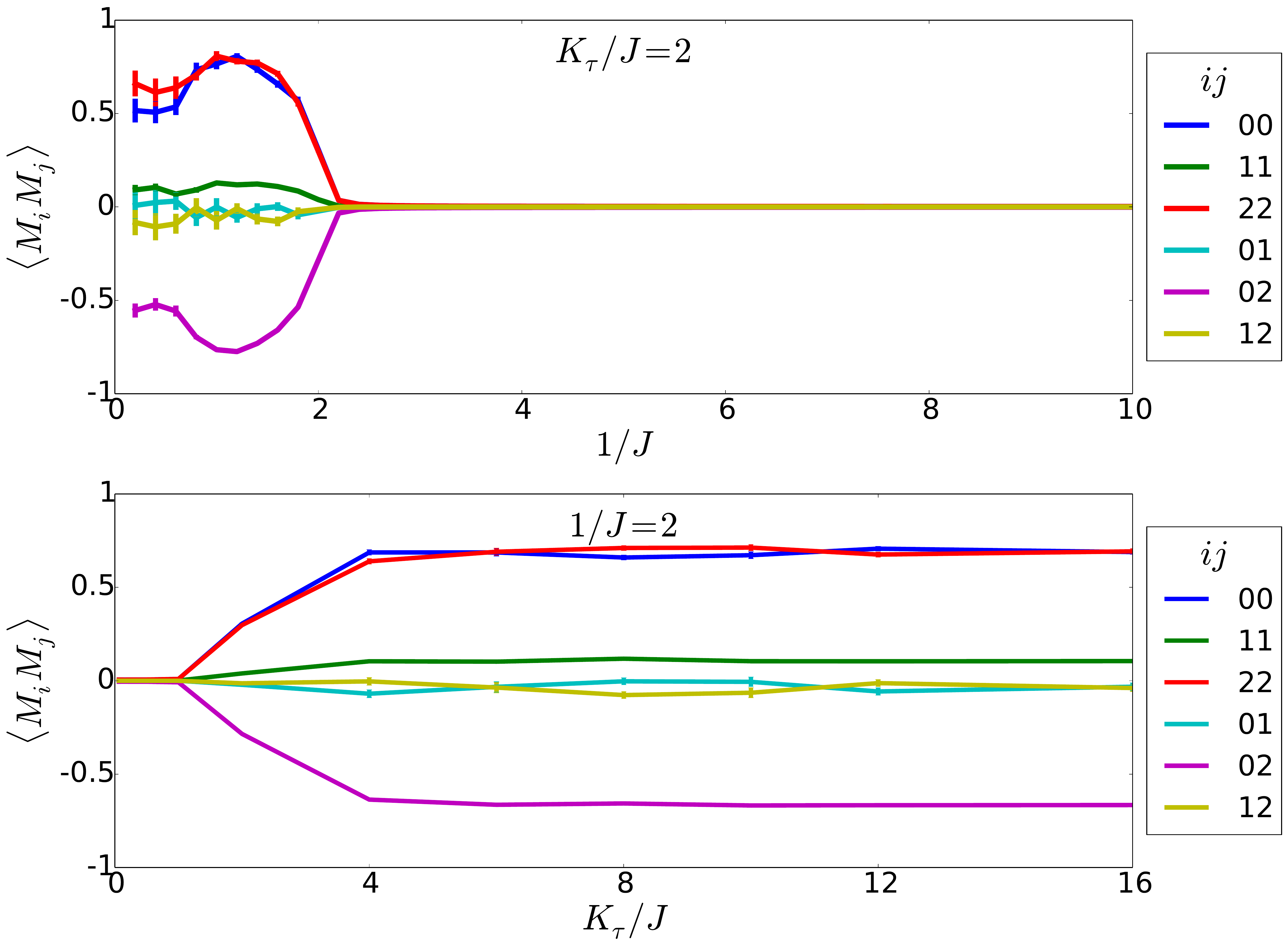}
\label{fig:mag}
}
\caption{(Color online) (a) Plots of the specific heat $c$ as a function of $1/J$ for a range of values of  $K_{\tau}/J$. (b) The sublattice magnetic correlations plotted along the lines $K_\tau/J=2$ and $1/J=2$. Both sublattices have an identical ordering as described in the text, and there are no magnetic correlations between the sublattices.}
\end{figure}

In Fig. \ref{fig:cpd}, we plot the phase diagram as a function of $1/ J$ vs. $K_\tau/J$, which is obtained from the peaks of specific heat (Fig. \ref{fig:cT}), and complemented by the calculation of selected critical couplings from the crossings of Binder ratios. We remind the reader that the paramagnetic phase of above model of Eq. \eqref{eq_oneham} corresponds to the  spin liquid phase of the microscopic Hamiltonian.

To explore the stability of the spin liquid to small Heisenberg perturbations we plot the specific heat against $1/J$ for a range of values of $K_{\tau}/J$ in Fig. \ref{fig:cT}. This  shows that the spin liquid is stable for small $J$, and that it remains so until a critical value is reached, with the sharp peak in the specific heat (and concomitant development of magnetic correlations) indicating a  phase transition to the magnetically ordered state. We remark that the broad feature of specific heat seen at large $1/J$  for large $K_\t$ in Fig. \ref{fig:cT}, which is absent in the $K_\t\to\infty$ limit \cite{PhysRevB.29.5250}, is due to the thermal excitation of the  link variables $\eta_{ij,\tau}$ which occurs at $1/J\sim K_{\tau}/J$.

\begin{figure}
\centering
\subfigure[]{
\includegraphics[width=\columnwidth]{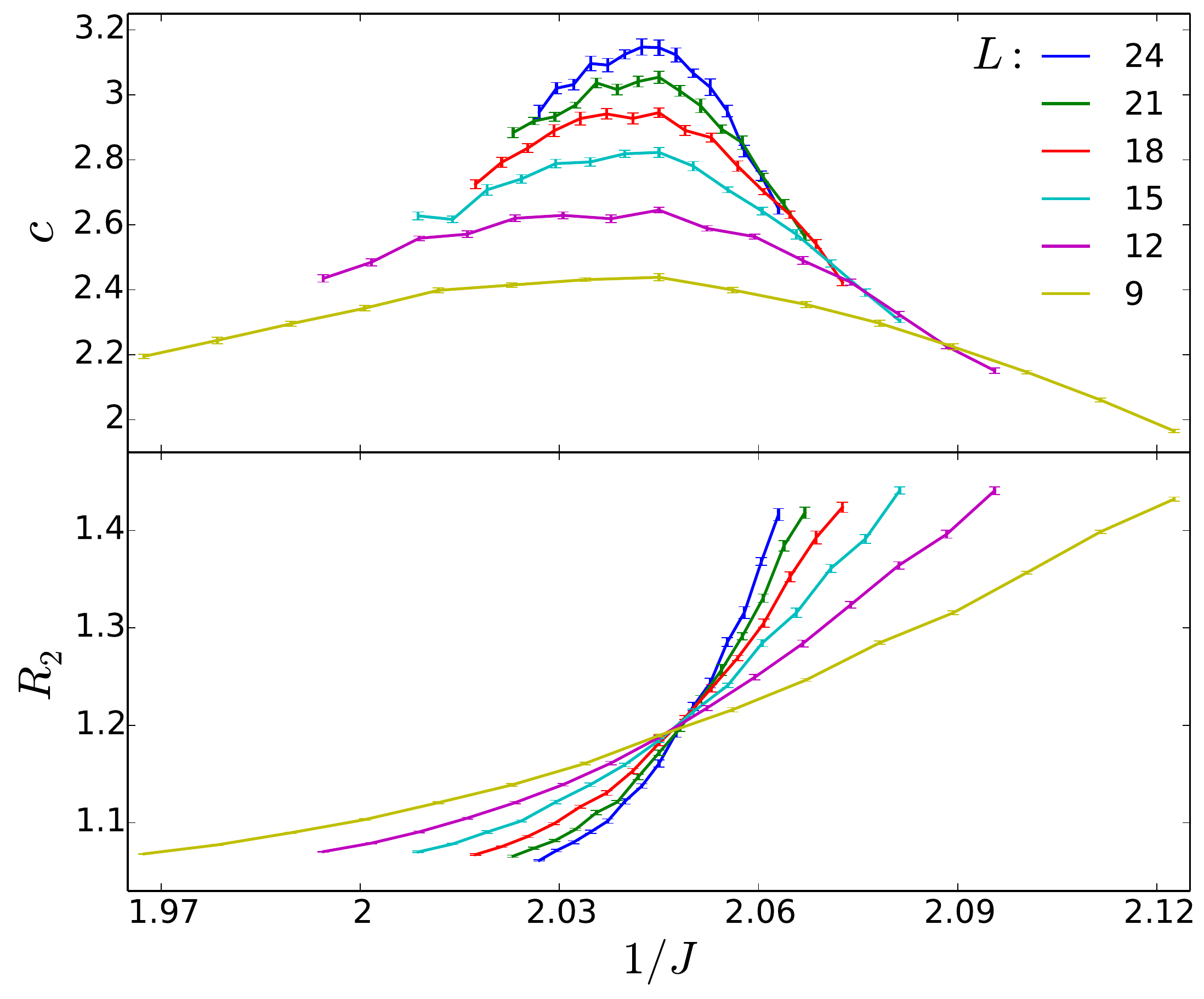} 
\label{fig_binder}
}
\subfigure[]{
\includegraphics[width=\columnwidth]{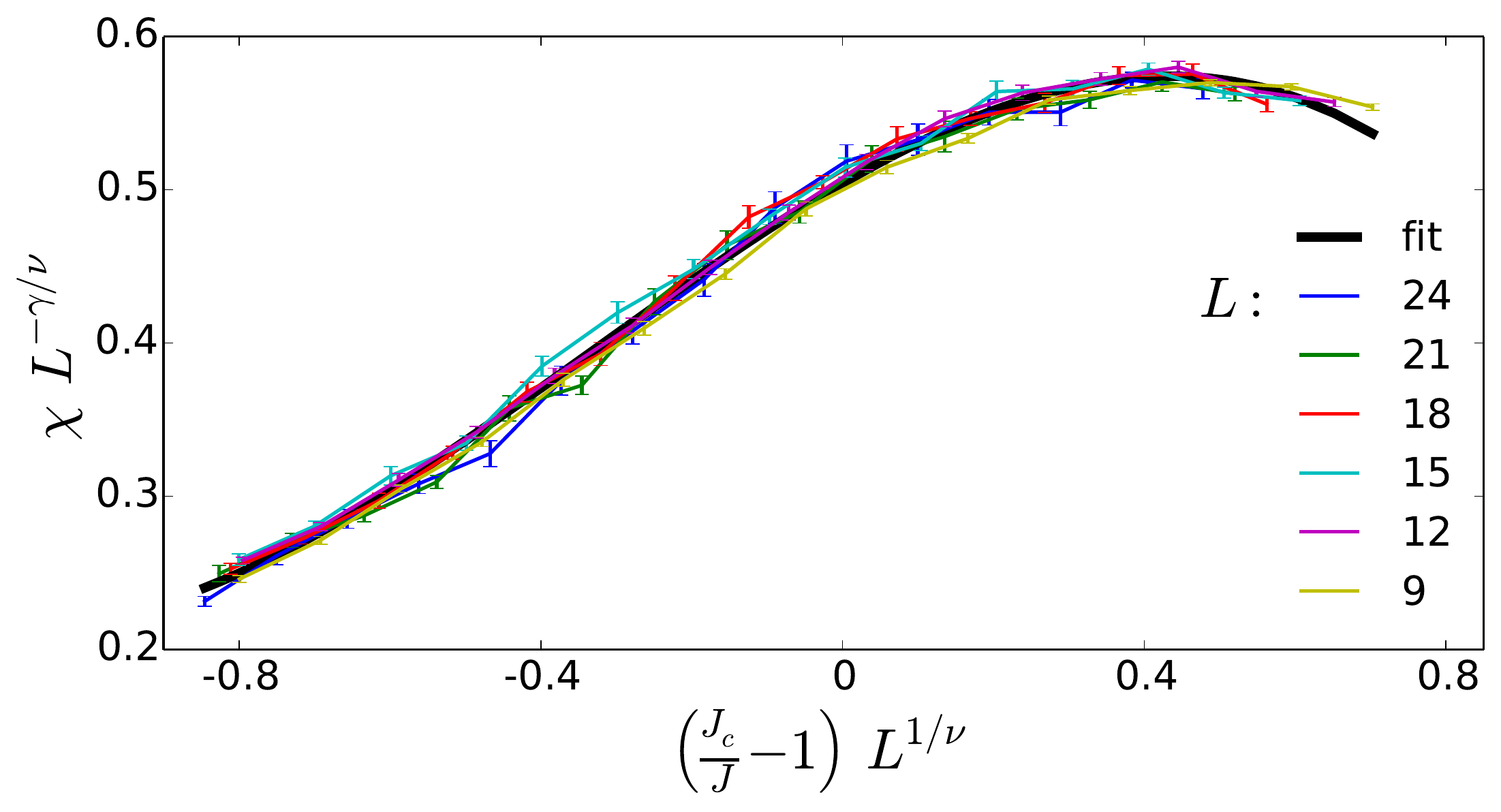} 
\label{fig_collapse}
}
\caption{(Color online) (a) The specific heat $c$ near the transition for $K_\tau/J=2$ and the Binder ratio $R_2$ gives the value of the critical coupling $1/J_c=2.0505$. (b) The scaling collapse of the magnetic susceptibility.}
\end{figure}

Below the critical values of $1/J$ the system is magnetically ordered as described before.  To investigate the nature of the ordering, we take a 3-site unit cell for each sublattice, and investigate the correlations within a sublattice, and between the two sublattices. The correlations within a sublattice are plotted in Fig.  \ref{fig:mag}  for two cuts of the $1/J$-$K_\t$  plane, given by $K_\t/J=2$ and $1/J=2$ respectively. 
For these the magnetisations are ordered as $M_1>M_2>M_3$ at each step of the Monte Carlo simulation, indicating the magnetic ordering $\langle M_2^2\rangle\simeq0$, $\langle M_1^2\rangle\simeq\langle M_3^2\rangle\simeq - \langle M_1 M_3\rangle$ displayed in Fig. \ref{fig:tri_order}. For the computations described in this section we take the order parameter to be $M=M_1-M_3$. We use this to calculate the Binder ratio 
$R_2 = \langle M^4\rangle/\langle M^2\rangle^2$ 
to determine the position of the phase transition to higher accuracy. This is shown in Fig. \ref{fig_binder}. We find no correlations between the respective orderings of the two sublattices, yielding the full magnetic ordering of Fig. \ref{fig_order_kag} in terms of the $\tau$ spins. From Fig.  \ref{fig:mag} it is clear that  the magnetic order does not change within the ordered region as far as we can resolve within our numerical calculations. 

To probe the critical point further, we study the critical exponents, which we extract via a scaling collapse of the magnetic susceptibility as shown in Fig. \ref{fig_collapse}.  We  restricted to small lattices, and focused on $L\times L\times L$ with $L=9,12,15,18,21,24$. These rather small sizes to which we are restricted lead to notable finite size effects. Within error however, we find that the scaling of the critical exponents for finite $K_\tau$ match those for $K_\tau=\infty$. In addition, for $K_\tau=\infty$ we went to $L=27,30$ and obtained results consistent with the $3D$-$XY$ critical exponents. While we do not completely understand why the present exponents are very close to those for $3D$-$XY$, to the best of our knowledge  the actual exponents for a $3D$-$XY\times Z_2$ critical point are not known. A related issue is whether one can separate the $3D$-$XY$ and the $Z_2$ transition to open up a phase in between where the $Z_2$ (inversion) symmetry is broken but where there is no magnetic order. We have not been able to achieve this in the present model, but this may be due to the fact that the $Z_2$ can be described by Ising variables sitting on the links of the medial honeycomb lattice, which is the Kagome lattice. This may cause frustration among the $Z_2$ variables and could prevent them from ordering alone in absence of the $XY$ field.

\paragraph*{The parameters of the generalized toric code and the microscopic model:} 
 we conclude with some comments on the relationship of the parameters of the generalized toric code Hamiltonian (Eq. \eqref{eq_gen_toric}) to the parameters of the microscopic model of Eq. \eqref{eq_toric}.  Firstly, the couplings $a_\bigtriangleup$ and $a_\bigtriangledown$ necessarily appear with a relative negative sign in Eq. \eqref{eq_toric} when $a_{\hexagon}>0$, whereas we consider $a_\bigtriangleup=a_\bigtriangledown$. The relative sign can be removed however by a transformation which rotates the $\mu$ spins about the $\mu^z$ axis on one of the triangular sublattices of the medial lattice such that $\mu^x\rightarrow-\mu^x$ for that sublattice. 
Next, we comment that the presence of the Heisenberg interaction would cause the coupling constants to change. These renormalizations would be small when $J_x,J_y>J_H$ (we assume $J_x,J_y>0$). 
Finally, we have done our calculation in the zero magnetic charge sector which is the relevant sector in the regime where $a_{\hexagon}$ is the dominating coupling constant of the generalized toric code model (eq. \ref{eq_gen_toric}). This  may not be so however as suggested from the couplings of the microscopic model. In the present case we assume that of the different magnetic charge sectors (which remain good quantum numbers in presence of the Heisenberg perturbations), the zero charge sector always remains the ground state. Numerical verification of this assumption forms a topic of future study.

\section{Effect of a magnetic field on the Kekul\'{e}-Kitaev model}
\label{sec_zeeman}

We now briefly discuss the effect of a magnetic field on the Kekul\'{e}-Kitaev model. We consider a Hamiltonian of the form
\begin{align}
H_Z=H_K-\mu\sum_i{\vec h}\cdot \vec{\sigma_i}
\end{align}
where ${\vec h}$ is the magnetic field. As a single spin operator creates two fluxes which cost energy, in the limit of small field the effect of the time-reversal symmetry breaking Zeeman term can be obtained by perturbation theory within the ground state (zero flux) sector.

\begin{figure}
\centering
\includegraphics[scale=0.3]{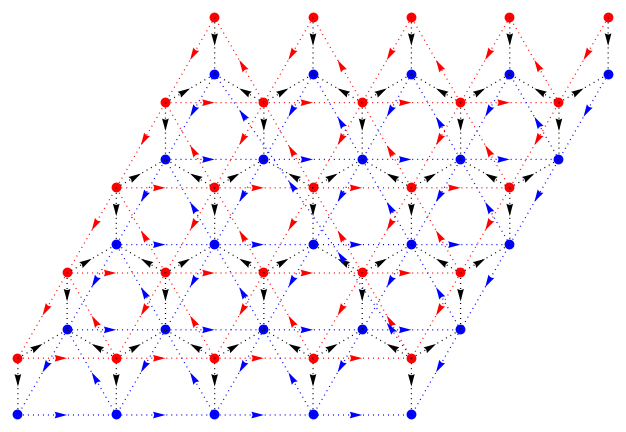}
\caption{(Color online) The nearest and the next nearest neighbour hopping for the isotropic model with two site unit cell. Going along the arrow, conncting two sites $i$ and $j$ gives a term $ic_ic_j$ in the tight binding Hamiltonian.}
\label{fig_zeeman}
\end{figure}

Let us thus describe the effect of the Zeeman term on the Majorana fermions.
The first non-trivial interaction that breaks time reversal symmetry is a three spin term, which when written in the zero flux sector provides next nearest neighbour hopping to the Majorana fermions (similar to the case in the original Kitaev model, other terms renormalize the nearest neighbour hopping or provide short range four fermion interactions which are irrelevant at the free Majorana fixed point). At the isotropic line ($J_x=J_y=J_z$), we can use a two site unit cell (used only in this section) for the Majorana fermions to make our results transparent. The hopping Hamiltonian obtained from including the Zeeman term through perturbation theory is shown in Fig. \ref{fig_zeeman}. It must be noticed that as one goes along the arrows in one hexagonal plaquette, the winding of the red and the blue arrows are mutually opposite contrary to the case in the usual Kitaev model.\cite{kitaev2006anyons} This means that the mass term obtained at the Dirac point does not invert from one valley to another and hence the Chern number of the two gapped bands are zero. This means that the state so obtained is not a chiral spin liquid as was obtained in the original Kitaev model\cite{kitaev2006anyons} and hence does not support gapless edge modes in this gapped time reversal symmetry broken QSL state.

\section{Summary and Discussions}
\label{sec_discuss}

Motivated by the idea to understand quantum spin liquid phases, and phase transitions from them, we have studied a variant of the Kitaev model with and without antiferromagnetic Heisenberg interactions. We found that the phase diagram even in the absence of perturbations  is quite different from the original Kitaev model, and includes regimes captured by toric code models on the Kagome lattice. In the presence of Heisenberg interactions, in the toric code limit, the system shows an interesting quantum phase transition to a magnetically ordered phase where the magnetic order itself is chosen through a  quantum `order by disorder' mechanism. Using a combination of field theory and Monte Carlo studies we have been able to study the phase transition which likely belongs to the $3D$-$XY\times Z_2$ universality class. Such a controlled description of a continuous quantum phase transition out of a QSL to a non-trivial magnetically ordered phase in a microscopic model is quite interesting in the context of recent interest in understanding novel QSL phases. We note that it has been claimed\cite{Japan} that a transition belonging to the $3D$-$XY\times Z_2$ universality class can be driven by fluctuations to a weak first order transition. However, we have not seen signatures of such a discontinuous transition in our numerics. While the possibility of a weak first order transition cannot be completely ruled out, we hope that future numerics on larger system sizes will be able to solve this issue.

The nature of the phases and phase transitions arising from Heisenberg antiferromagnetic perturbations in the limit of isotropic Kitaev couplings appears to be more complicated, due to the absence of a special $SU(2)$ invariant point\cite{PhysRevLett.105.027204} in the present case. While from general considerations the QSL phase is expected to be stable to Heisenberg perturbations, a determination of the window of stability and the resultant phase to which this QSL gives way constitutes an interesting avenue of future study.

\section{Acknowledgements}
We thank V. Jouffrey, A. Paramekanti, Y. B. Kim, F. Pollmann and K. P. Schmidt for discussions. 

\appendix
\section{The details of the lattice and Brillouin zone}
\label{appen_bz}
\begin{figure}[htp!]
\centering
\includegraphics[scale=0.2]{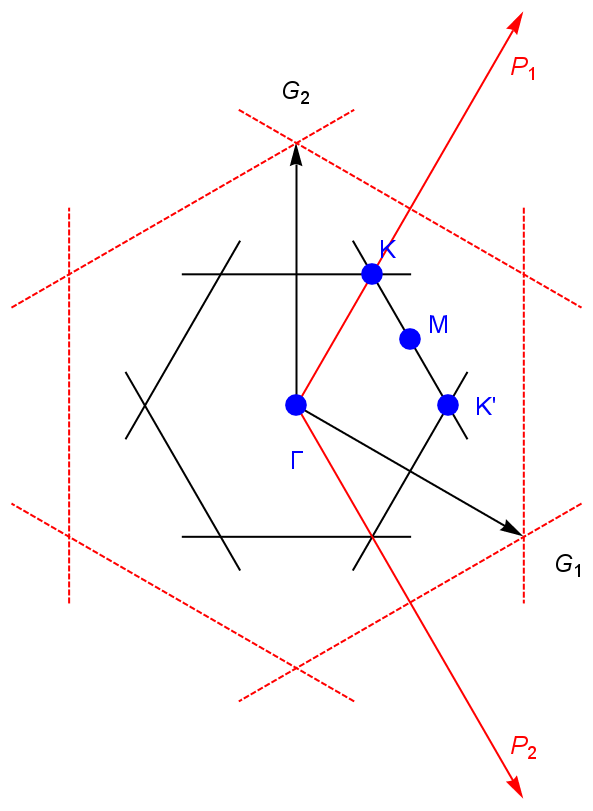}
\caption{(Color online) The Brillouin zones: The red (black) lines and arrows denote the Brillouin zone of the isotropic (anisotropic) honeycomb lattice with two (six) point unit cell. The Reciprocal vectors are denoted by $P_1 (G_1)$ and $P_2(G_2)$ for the two (six) site cases. ${\bf G}_1=\left\{\frac{2 \pi }{3},-\frac{2 \pi }{3\sqrt{3}}\right\};~~~{\bf G}_2=\left\{0,\frac{4 \pi }{3\sqrt{3}}\right\}$. We note that the zone corners of the two-site unit cell lattice maps to the zone centre of the six-site unit cell case.}
\label{fig_bz_rel}
\end{figure}

The lattice vectors, as shown in Fig. \ref{fig_rotated}, are given in the caption of the same figure. The Brillouin zone is presented in Fig. \ref{fig_bz_rel}, along with the reciprocal lattice vectors.

\section{The exact solution of the Kekul\'e-Kitaev model and the applicability of Lieb's theorem}
\label{appen_lieb}
\begin{figure}
\centering
\includegraphics[scale=0.45,angle=-90]{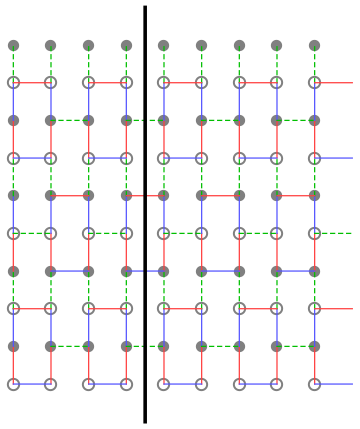}
\caption{(Color online) The Kekul\'e-Kitaev model on a brick lattice, which is a square lattice with selectively deleted bonds. The reflection symmetry about the black line allows the application of Lieb's theorem.}
\label{fig_brick}
\end{figure}

\begin{figure*}
\centering
\subfigure[]{
\includegraphics[scale=0.23]{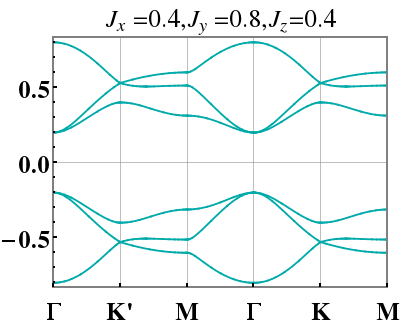}
}
\subfigure[]{
\includegraphics[scale=0.23]{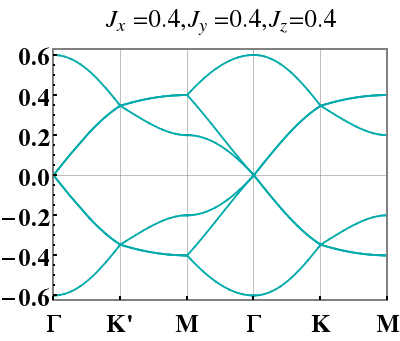}
}
\subfigure[]{
\includegraphics[scale=0.23]{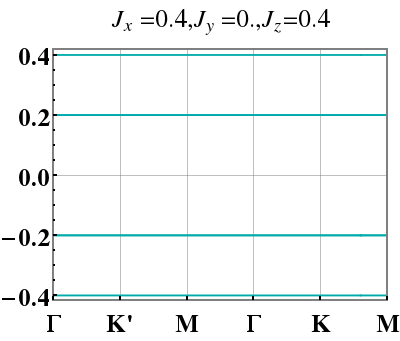}
}
\subfigure[]{
\includegraphics[scale=0.23]{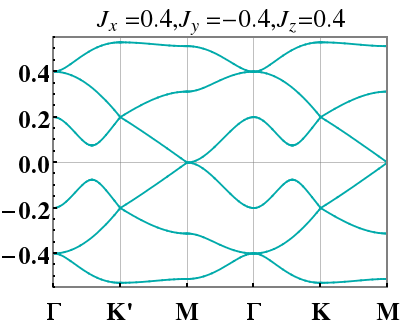}
}
\subfigure[]{
\includegraphics[scale=0.23]{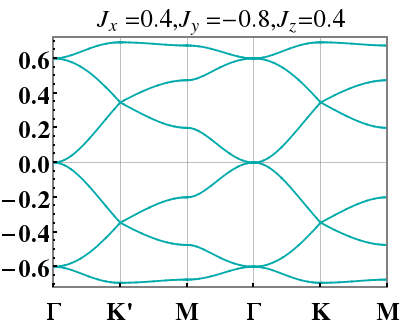}
}
\subfigure[$\Gamma$]{
\includegraphics[scale=0.23]{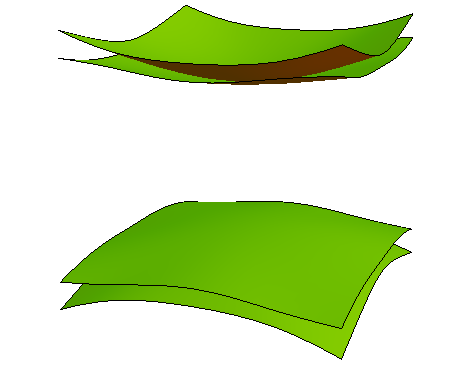}
}
\subfigure[$\Gamma$ (doubly degenrate)]{
\includegraphics[scale=0.23]{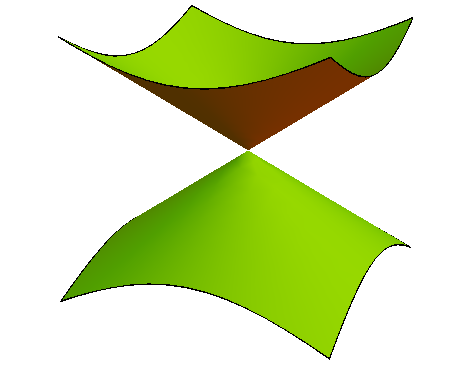}
}
\subfigure[$\Gamma$ (doubly degenrate)]{
\includegraphics[scale=0.23]{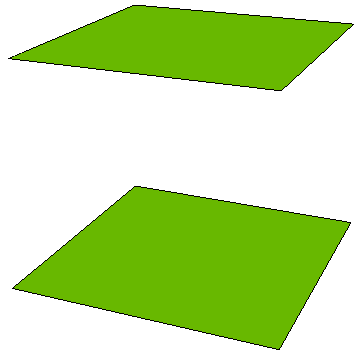}
}
\subfigure[$M$]{
\includegraphics[scale=0.23]{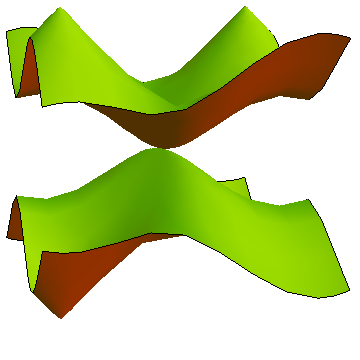}
}
\subfigure[$\Gamma$]{
\includegraphics[scale=0.23]{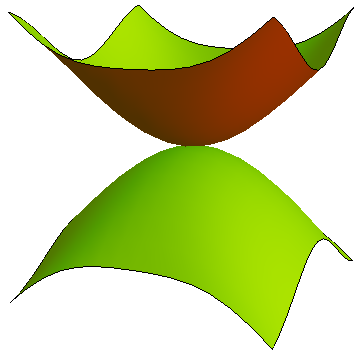}
}
\caption{(Color online) The Majorana band structure: (a)-(e) denotes the cut shown in Fig. \ref{fig_pd}. (f)-(j) denotes the low energy section of the bands corresponding to (a)-(e) respectively.}
\label{fig_band}
\end{figure*}

To obtain the exact solution, following Kitaev, we can define the spins in terms of four  mutually anticommuting Majorana fermions
\begin{align}
\sigma^\alpha_i=ib^\alpha_ic_i
\end{align}
subject to the constraint
\begin{align}
D_i\equiv b^x_ib^y_ib^z_ic_i=1\,,~~~~~~~~~~\forall i\,.
\end{align}
The $D_i$ commute with the Hamiltonian, $[H,D_i]=0$,  resulting in a $Z_2$ gauge structure, for which the $D_i$ generate the $Z_2$ gauge transformation. Continuing a similar treatment as in the original case we write for the $\alpha$-th link $\sigma^\alpha_i\sigma_j^\alpha=(ib_i^\alpha c_i)(ib_j^\alpha c_j)=-i\hat{u}^\alpha_{ij}c_ic_j$, where, $\hat{u}^\alpha_{ij}=ib_i^\alpha b_j^\alpha=-\hat{u}^\alpha_{ji}$. All the $\hat{u}^\alpha_{ij}$ commute amongst themselves as well as with the Hamiltonian, and are thus constants of motion. However they are not however gauge invariant. Rather they are the gauge potentials of the $Z_2$ gauge theory. The plaquette operators of Eq. \eqref{plaq_ops} take the form
\begin{align}
\mathcal{W}_\alpha(P)=\prod_{ij\in\beta-link,\in P}\hat{u}_{ij}^\beta~~~~~~(\beta\neq\alpha;~~ i\in odd, j\in even)
\end{align}
and the Hamiltonian becomes 
\begin{align}
\mathcal{H}=&i\sum_{\alpha-\rm links}J_\alpha\hat{u}_{ij}^\alpha c_ic_j.
\end{align}
We can replace the $\hat{u}_{ij}^\alpha$ with their eigenvalues $\pm 1$, recasting the systems as a tight-binding model where the $c$ Majorana femions hop on a bipartite structure such that the hopping amplitudes $t_{ij}\neq 0$ only if $i\in~odd$ and $j\in~even$, and all loops contain an even number of edges and are planar. Keeping intact the relevant symmetries of the model, the lattice can be deformed to the one shown in Fig. \ref{fig_brick}. 
Clearly, the figure shows that the results of the Lieb's theorem\cite{PhysRevLett.73.2158} are applicable in this case and hence the ground state lies in the zero flux sector for the Majorana fermions. This flux sector is ensured by choosing $u^\beta_{ij}=+1$ when $i\in odd$ and $j\in even$. 
Let us note an important difference with the usual Kitaev model. As each plaquette contains a definite kind of bond three times or not at all, a global sign change in any of $J_x,J_y$ or $J_z$ can no longer be absorbed in changing the sign of the corresponding $u^\beta_{ij}$ (through a gauge transformation).  As a result the ground state energy in the sectors $(J_x,J_y,J_z)$ and $(J_x,J_y,-J_z)$ will be different. 
The band structure of the Majorana fermions is obtained by diagonalizing the free Majorana Hamiltonian, giving rise to 6 particle-hole symmetric Majorana bands. We plot them and analyze the band structure in different parameter regimes to find the phase diagram  shown in Fig. \ref{fig_pd}. Some representative band structures are shown in Fig. \ref{fig_band}.
\subsection*{Generation of mass around the isotropic line $(J_x=J_y=J_z)$}
On the isotropic line, we have four bands, with linear dispersion, touching at the $\Gamma$ point. The low energy $k.p$ Hamiltonian for these four bands (for $c$ majorana fermions) can be obtained (for the zero flux sector) from the lattice hopping Hamiltonian by expanding about the $\Gamma$ point to the linear order and projecting to the four bands. This has the form
\begin{align}
\mathcal{H}_{\rm iso}=\frac{3}{4}\left[\begin{array}{cccc}
0 & 0 & e^{i\tau_1}k_x  & e^{i\tau_2}k_y \\
0 & 0 & e^{i\tau_3}k_y  & e^{i\tau_4}k_x \\
e^{-i\tau_1}k_x  & e^{-i\tau_3}k_y  & 0 & 0\\
e^{-i\tau_2}k_y  & e^{-i\tau_4}k_x  & 0 & 0\\  
\end{array}\right]
\end{align}
where (we have numerically calculated) $\tau_1=0.871499\pi,\tau_2=-0.138271\pi, \tau_3=-0.0833333\pi, \tau_4=-0.0931037\pi$. The doubly degenerate eigenvalues are given by $\xi_\pm({\bf k})=\pm\frac{3}{4}|{\bf k}|$. This is the linearly dispersing band touching shown in Fig. \ref{fig_band}.

We now wish to move away from the isotropic line by adding an anisotropy of the following form: $J_x=J+\delta, J_y=J_z=J$. This anisotropy pattern is similar to the Kekul\'e order parameter of Ref. \onlinecite{PhysRevB.82.035429} (with $\alpha=0$ in their notation in Eqs. 9 and 10 of that paper). The mass matrix has the form:
\begin{align}
\mathcal{H}_{\rm mass}=\frac{\delta}{4}\left[\begin{array}{cccc}
0 & 0 & e^{i\eta_1}  & \sqrt{3}e^{i\eta_2} \\
0 & 0 & \sqrt{3}e^{i\eta_3}  & e^{i\eta_4} \\
e^{-i\eta_1}  & \sqrt{3}e^{-i\eta_3}  & 0 & 0\\
\sqrt{3}e^{-i\eta_2}  & e^{-i\eta_4}  & 0 & 0\\  
\end{array}\right]
\end{align}
where $\eta_1=0.371499\pi, \eta_2=0.361729\pi, \eta_3=-0.583333\pi, \eta_4=0.406896\pi$. The corresponding doubly degenerate bands have the dispersions, 
\begin{align}
\xi_\pm({\bf k})=\pm\sqrt{\frac{9}{16}{\bf k}^2+\frac{1}{4}\delta^2}\,.
\end{align}
The double degeneracy is lifted away from the Dirac point due to higher order terms as seen in the Fig. \ref{fig_band}.

\section{Perturbative derivation of the toric code model}
\label{appen_toric_pert}

The toric code Hamiltonian $H_{TC}$ of Eq. \eqref{eq_toric} appears through degenerate perturbation theory on restricting the Hilbert space of the full model to the ground state manifold of  the bond doublets $|\Uparrow\rangle$, $|\Downarrow\rangle$ of Eq. \eqref{bond_doublets} at the $z$-bonds of the honeycomb lattice (which form a Kagome lattice). The Hamiltonian is computed order by order in $1/J_z$:\cite{kitaev2006anyons}
\begin{align}
H_{eff}=P\left[V+VG_0'V+VG_0'VG_0'V+\cdots\right]P
\end{align}
where $G_0'=(1-P)\frac{1}{E-H_0}(1-P)$ with $P=\prod_K\left(|\Uparrow\rangle_K {}_K\langle\Uparrow|+|\Downarrow\rangle_K {}_K\langle\Downarrow|\right)$ being the projector onto the ground state manifold ($K$ labels the sites of the Kagome lattice), and 
\begin{align}
H_0&=-J_z\sum_{\rm z-links}\sigma_i^z\sigma_j^z\,,\\
V&=-J_y\sum_{\rm y-links}\sigma_i^y\sigma_j^y-J_x\sum_{\rm x-links}\sigma_i^x\sigma_j^x\,.
\end{align}
The first non-trivial terms occur at third order
\be
\begin{aligned}
H_{eff}^{(3)}=&-\frac{{ 3} J_x^3}{8J_z^2}\sum_{\bigtriangleup,\{I,J,K\in\bigtriangleup\}}\tau^x_I\tau^x_J\tau^x_L\\
&+\frac{{ 3}  J_y^3}{8J_z^2}\sum_{\bigtriangledown,\{I,J,K\in\bigtriangledown\}}\tau^x_I\tau^x_J\tau^x_L\,\\
=&-\frac{{ 3} J_x^3}{8J_z^2}\sum_{\bigtriangleup} \lF+\frac{{ 3}  J_y^3}{8J_z^2}\sum_{\bigtriangledown} \rF\,,
\end{aligned}
\ee
while next non-trivial contribution to the effective Hamiltonian occurs at sixth order. The second and fourth order terms are constant, while the fifth order term gives a subleading correction to the third order term. At sixth order there is a new term $\tau^z_{I_1}\tau^z_{I_2} \tau^z_{I_3} \tau^z_{I_4} \tau^z_{I_5} \tau^z_{I_6} $ 
associated with each hexagon of the Kagome lattice, where $I_j$ labels the $z$-bonds at the vertices of the hexagon. In addition there is a contribution from pairs of neighbouring $\bigtriangleup$, $\bigtriangledown$ triangles, $\tau^x_{J_1}\tau^x_{J_2} \tau^x_{J_3} \tau^x_{K_1} \tau^x_{K_2} \tau^x_{K_3}$ where $J_j$, $K_j$ label the $z$-bonds at the vertices of the two respective triangles, one of the bonds being common to both. Counting up these two contributions gives the relevant sixth order term of the effective Hamiltonian
\begin{align}
H_{eff}^{(6)}=-\frac{ 3 J_x^3 J_y^3}{256J_z^5} \sum_{\hexagon} \hF
+  \frac{ 7 J_x^3 J_y^3}{64 J_z^5}\sum_{\langle\bigtriangleup,\bigtriangledown \rangle } \lF \rF\,,
 \end{align}
 where $\hF = \prod_{K\in \hexagon} \t^z_K$ and $\langle,\rangle$ indicates the sum is to be taken over neighbouring pairs of triangles. The operators $\hF$ are  equivalent to the $\mathcal{W}_z$ plaquette operators of the full model, while the operators $\lF$ and $\rF$ are  equivalent to the $\mathcal{W}_y$ and $\mathcal{W}_x$ plaquette operators respectively, and so  $H_{TC} = H_{eff}^{(3)} +H_{eff}^{(6)}$ is a good effective model, it captures the leading physics of all degrees of freedom in  the strong bond limit.

\section{The soft spin Landau-Ginzburg action}
\label{appen_ising}
The critical action for the triangular lattice can be derived following the work of  Blanckstein et. al.\cite{PhysRevB.29.5250} Taking the unit vectors of the triangular lattice 
\begin{align}
{\bf a'}_1=[1,0];~~{\bf a'}_2=[1,\sqrt{3}]/2;
\end{align}
and the reciprocal lattice vectors: 
\begin{align}
{\bf G'}_1=2\pi[1,1/\sqrt{3}];~~{\bf G'}_2=2\pi[0,2/\sqrt{3}],
\end{align}
the soft modes occur at 
\begin{align}
{\bf K}_\pm=\pm[4\pi/3,0]\,.
\end{align}
The soft mode expansion of the spin around these momenta is then given by Eq. \ref{eq_soft_amp}.
The generators of the symmetries (of the Hamiltonian) on the triangular lattice are: (1) unit translation along ${\bf a}_1$, (2) unit translation along ${\bf a}_2$, (3) global $Z_2$.

The two amplitudes transform as:
\begin{widetext}
\begin{align}
\mathcal{T}_x: \{x,y\}\rightarrow\{x+1,y\}:\mu^z({\bf r})\rightarrow \mu^z({\bf r+a_1}):&\Rightarrow
\{\psi_+,\psi_-\}\rightarrow\{\psi_+e^{-i4\pi/3},\psi_-e^{i4\pi/3}\}\\
\mathcal{T}_y: \{x,y\}\rightarrow\{x,y+1\}:\mu^z({\bf r})\rightarrow \mu^z({\bf r+a_2}):&\Rightarrow
\{\psi_+,\psi_-\}\rightarrow\{\psi_+e^{-i2\pi/3},\psi_-e^{i2\pi/3}\}\\
Z_2:\{x,y\}\rightarrow\{x,y\}:\mu^z({\bf r})\rightarrow -\mu^z({\bf r}):&\Rightarrow \{\psi_+,\psi_-\}\rightarrow\{-\psi_+,-\psi_-\}\\
C_6:\{x,y\}\rightarrow \{-y,x+y\}:\mu^z({\bf r})\rightarrow \mu^z({\bf r'}):&\Rightarrow \{\psi_+,\psi_-\}\rightarrow\{\psi_-,\psi_+\}\\
\mathcal{I}:\{x,y\}\rightarrow\{-x,-y\}:\mu^z({\bf r})\rightarrow \mu^z({\bf -r}):&\Rightarrow \{\psi_+,\psi_-\}\rightarrow\{\psi_-,\psi_+\}
\end{align}

\end{widetext}
The resultant Euclidean Landau-Ginzburg action is given by Eq. \ref{eq_zeroLGW}.

For our case we have two copies of the triangular lattice, which comprise the honeycomb lattice, and the two triangular lattices transform into each other under inversion about the midpoint of the bond joining the sublattices of the medial honeycomb lattice. Also the inversion about the bond ($\mathcal{I}$) is no longer there and the $C_6$ symmetry is replaced by a $C_3$ symmetry and finally there is reflection about the vertical bond ($\mathcal{R}$). Under these new symmetries the transformation of the soft modes are:
\begin{widetext}
\begin{align}
C_3:\{x,y\}\rightarrow\{-x-y,x\}:\mu^z(1,{\bf r})\rightarrow \mu^z(1,{\bf r'}):&\Rightarrow\{\psi^{(1)}_+,\psi^{(1)}_-\}\rightarrow\{\psi^{(1)}_+,\psi^{(1)}_-\};\nonumber\\
\mathcal{I}_B:\{x,y\}\rightarrow\{-x,-y\}:\mu^z(1,{\bf r})\rightarrow \mu^z(2,-{\bf r}):&\Rightarrow\{\psi^{(1)}_+,\psi^{(1)}_-\}\rightarrow\{\psi^{(2)}_-,\psi^{(2)}_+\};\nonumber\\
\mathcal{R}:\{x,y\}\rightarrow\{-x-y,y\}:\mu^z(1,{\bf r})\rightarrow \mu^z(1,{\bf r'}):&\Rightarrow\{\psi^{(1)}_+,\psi^{(1)}_-\}\rightarrow\{\psi^{(1)}_-,\psi^{(1)}_+\}
\end{align} 
\end{widetext}
This gives the action given by Eq. \ref{eq_LGW}.

\section{Classical action on stacked honeycomb lattice}
\label{appen_class}

The Hamiltonian in Eq. \ref{eq_oneham} has the following form:
\begin{align}
H=a\sum_{\langle\langle ik\rangle\rangle}\mu^z_i\mu^z_{k}-\Gamma\sum_i\mu^x_i-\Lambda\sum_{\langle ij\rangle}\mu^x_i\mu^x_j
\end{align}

To obtain the Euclidean action we resolve the partition function 
$\mathcal{Z}=\sum_{\{\mu^z\}}\langle\{\mu^z\}|e^{-\beta H}|\{\mu^z\}\rangle$ 
in the time direction by inserting compete basis of states at each time slice 
\begin{align}
\mathcal{Z}=\sum_{\{\mu^z_0\}}\sum_{\{\mu^z_1\}}\cdots\sum_{\{\mu^z_\tau\}}\cdots\sum_{\{\mu^z_N\}}\mathcal{M}_{0,1}\cdots\mathcal{M}_{\tau,\tau+1}\cdots\mathcal{M}_{N,0}
\end{align}
where $\mathcal{M}_{\tau,\tau+1}=\langle\{\mu^z_\tau\}|e^{-\Delta\tau H}|\{\mu^z_{\tau+1}\}\rangle$.
\begin{widetext}
To simplify this we first rewrite
\begin{align}
\mathcal{M}_{\tau,\tau+1}&=e^{-\Delta\tau a\sum_{\langle\langle ik\rangle\rangle}\mu^z_{i\tau}\mu^z_{k\tau}}\langle\{\mu^z_\tau\}|e^{\Delta\tau \Lambda\sum_{\langle ij\rangle}\mu^x_i\mu^x_j}e^{\Delta\tau \Gamma\sum_i\mu^x_i}|\{\mu^z_{\tau+1}\}\rangle\nonumber\\
&\propto e^{-\Delta\tau a\sum_{\langle\langle ik\rangle\rangle}\mu^z_{i\tau}\mu^z_{k\tau}}\langle\{\mu^z_\tau\}|\left[\prod_{\langle ij\rangle}\left(1+e^{-2K_\tau}e^{i\pi\left[\frac{1-\mu^x_i}{2}+\frac{1-\mu^x_j}{2}\right]}\right)\right]e^{\Delta\tau \Gamma\sum_i\mu^x_i}|\{\mu^z_{\tau+1}\}\rangle\nonumber\\
&\propto e^{-\Delta\tau a\sum_{\langle\langle ik\rangle\rangle}\mu^z_{i\tau}\mu^z_{k\tau}}\sum_{\{n_{ij,\tau}=0,1\}}\langle\{\mu^z_\tau\}|e^{\sum_{\langle ij\rangle}n_{ij,\tau}\left[-2K_\tau+i\pi\left[\frac{1-\mu^x_i}{2}+\frac{1-\mu^x_j}{2}\right]\right]}e^{-2\Delta\tau \Gamma\sum_i\frac{1-\mu^x_i}{2}}|\{\mu^z_{\tau+1}\}\rangle\nonumber\,.
\end{align}
where $\tanh(\Delta\tau \Lambda)=e^{-2K_\tau}$, $\Delta\tau=\beta/N$. Inserting a basis of $\mu^x$, and using the identity $\langle \mu^z|\mu^x\rangle=e^{i\pi\frac{1-\mu^x}{2}\frac{1-\mu^z}{2}}$, this becomes
\begin{align}
\mathcal{M}_{\tau,\tau+1}&\propto e^{-\Delta\tau a\sum_{\langle\langle ik\rangle\rangle}\mu^z_{i\tau}\mu^z_{k\tau}}\sum_{\{n_{ij,\tau}\}}\sum_{\{\mu^x_\tau\}}e^{\sum_{\langle ij\rangle}n_{ij,\tau}\left[-2K_\tau+i\pi\left[\frac{1-\mu^x_{i\tau}}{2}+\frac{1-\mu^x_{j\tau}}{2}\right]\right]}e^{-2\Delta\tau \Gamma\sum_i\frac{1-\mu^x_{i\tau}}{2}}e^{i\pi\sum_i\frac{1-\mu^x_{i\tau}}{2}\left[\frac{1-\mu^z_{i,\tau}}{2}+\frac{1-\mu^z_{i,\tau+1}}{2}\right]}\,.\nonumber
\end{align}
Gathering terms as
\begin{align}
\mathcal{M}_{\tau,\tau+1}&\propto e^{-\Delta\tau a\sum_{\langle\langle ik\rangle\rangle}\mu^z_{i\tau}\mu^z_{k\tau}} \sum_{\{n_{ij,\tau}\}}e^{-2K_\tau\sum_{\langle ij\rangle}n_{ij,\tau}}\sum_{\{\mu^x_\tau\}}e^{\sum_i\frac{1-\mu^x_{i\tau}}{2}\left[-2\Delta\tau \Gamma+i\pi\left[\frac{1-\mu^z_{i,\tau}}{2}+\frac{1-\mu^z_{i,\tau+1}}{2}\right]+i\pi\sum_jn_{ij,\tau}\right]}
\end{align}
the sum over $\{\mu^x_{\tau}\}$ gives
\begin{align}
\mathcal{M}_{\tau,\tau+1}&\propto e^{-\Delta\tau a\sum_{\langle\langle ik\rangle\rangle}\mu^z_{i\tau}\mu^z_{k\tau}}\sum_{\{n_{ij,\tau}\}}e^{-2K_\tau\sum_{\langle ij\rangle}n_{ij,\tau}}\prod_i\left(1+e^{-2\Delta\tau \Gamma}(-1)^{\left[\frac{1-\mu^z_{i,\tau}}{2}+\frac{1-\mu^z_{i,\tau+1}}{2}\right]+\sum_jn_{ij,\tau}}\right)\,.\nonumber
\end{align}
Introducing link variables $\eta_{ij,\tau}=\pm 1$ on space-like links through  $n_{ij,\tau}=\frac{1-\eta_{ij,\tau}}{2}$, this becomes
\begin{align}
\mathcal{M}_{\tau,\tau+1}&\propto e^{-\Delta\tau a\sum_{\langle\langle ik\rangle\rangle}\mu^z_{i\tau}\mu^z_{k\tau}} \sum_{\{\eta_{ij,\tau}\}}e^{K_\tau\sum_{\langle ij\rangle}\eta_{ij,\tau}}\prod_i\left(1+e^{-2\Delta\tau \Gamma}\left[\left(\prod_j\eta_{ij,\tau}\right)\mu^z_{i,\tau}\mu^z_{i,\tau+1}\right]\right)\,.\nonumber
\end{align}
Now writing $\tanh(\tilde{K})=e^{-2\Delta\tau\Gamma}$, we get
\begin{align}
\mathcal{M}_{\tau,\tau+1}&\propto e^{-\Delta\tau a\sum_{\langle\langle ik\rangle\rangle}\mu^z_{i\tau}\mu^z_{k\tau}} \sum_{\{\eta_{ij,\tau}\}}e^{K_\tau\sum_{\langle ij\rangle}\eta_{ij,\tau}}\prod_i\left(\cosh(\tilde{K})+\sinh(\tilde{K})\left[\left(\prod_j\eta_{ij,\tau}\right)\mu^z_{i,\tau}\mu^z_{i,\tau+1}\right]\right)\nonumber\\
&\propto e^{-\Delta\tau a\sum_{\langle\langle ik\rangle\rangle}\mu^z_{i\tau}\mu^z_{k\tau}} \sum_{\{\eta_{ij,\tau}\}}e^{K_\tau\sum_{\langle ij\rangle}\eta_{ij,\tau}+\tilde{K}\sum_i\left(\prod_j\eta_{ij,\tau}\right)\mu^z_{i,\tau}\mu^z_{i,\tau+1}}\,.
\end{align}
Finally gathering all terms we  write the partition function as
\begin{align}
\mathcal{Z}\propto\sum_{\{\mu^z_{k\tau}\}}\sum_{\{\eta_{ij,\tau}\}}e^{- J E}
\end{align}
where $E$ is the classical Hamiltonian which takes the form
\begin{align}\la{clmod}
E=\sum_{\langle\langle ik\rangle\rangle,\tau} \mu^z_{i\tau}\mu^z_{k\tau}-\frac{\tilde{K}}{ J}\sum_{i,\tau}\left(\prod_j\eta_{ij,\tau}\right)\mu^z_{i,\tau}\mu^z_{i,\tau+1}-\frac{K_\tau}{ J}\sum_{\langle ij\rangle,\tau}\eta_{ij,\tau}
\end{align}
where $ J=\Delta\tau a$.
\end{widetext}

\bibliography{biblio}
\end{document}